\documentclass[11pt]{article}

\usepackage{amsmath}
\usepackage{amssymb}
\usepackage{amsbsy}
\usepackage{amsfonts}
\usepackage{amsopn}
\usepackage{amstext}
\usepackage{graphicx}
\usepackage{amssymb}
\usepackage{amsfonts}
\usepackage{amsmath}
\usepackage{graphicx}
\usepackage[english]{babel}
\usepackage{color}
\usepackage[dvips]{epsfig}
\usepackage[dvips]{graphicx}
\usepackage{float}
\usepackage{units}
\usepackage{textcomp}
\usepackage{babel}

\usepackage{hyperref}
\usepackage{slashed}

\usepackage{graphicx} 
\usepackage{subfigure} 

\begin{document}

\setcounter{page}{1}

\pagestyle{plain} \vspace{1cm}

\begin{center}
\Large{\bf Localization of massive and massless fermion on two field brane }\\
\small \vspace{1cm} {\bf A. Farokhtabar\footnote{farokhtabarali@yahoo.com}}\quad
and\quad{\bf A. Tofighi \footnote{A. Tofighi@umz.ac.ir}} \\
\vspace{0.5cm} Department of Physics, Faculty of Basic Science,
University of Mazandaran, P. O. Box 47416-1467, Babolsar, Iran.
\end{center}

\date{\today}

\begin{abstract}
In this paper we study fermion localization and resonances on a
special type of braneworld model supporting brane splitting. In such models one
can construct multi-wall branes which cause considerable
simplification in field equations. We use a polynomial
superpotential to construct this brane. The suitable Yukawa coupling
between the background scalar field and localized fermion is determined.
The massive fermion resonance spectrum is obtained.
 The number of resonances is increased for higher values of Yukawa coupling. \\
{\bf PACS}: 
\end{abstract}
\newpage

\section{Introduction}
\label{sec:intro} Braneworld scenarios have attracted considerable
attention in the literature during  the last two decades because
theses models can
 address some important issues in
  theoretical physics problems such as hierarchy \cite{Randall:1999ee,ArkaniHamed:1998rs,Antoniadis:1998ig}
  and cosmological constant problem \cite{Rubakov:1983,Csaki:2000wz}.
  The branes in Randall and Sundrum (RS) models are fixed in some points along extra dimension  and have
  a $\delta$ function form \cite{Randall:1999ee,Randall:1999vf}. This brane-world model is very ideal and
  its formation has no dynamical mechanism.  But for realistic models thickness of brane should be considered.
  By now several thick brane construction mechanisms have been developed such as thick branes generated
   from pure gravity \cite{Arias:2002ew}-\cite{Liu:2009dt}, fermion self
   interaction branes \cite{Andrianov:2003hx,Andrianov:2005hm} and thick brane
   scenarios with the gravity coupled to five-dimensional scalar fields \cite{Bonjour:1999kz}-\cite{Kobayashi:2001jd}.
   In the last scenario, the scalar field configuration is usually a kink.
   It is found that a single kink becomes unstable when it moves in a discrete lattice with a large velocity
    while multi-kink solutions remain stable \cite{peyrard:1984}.
    This phenomenon is associated with interaction between kink and radiation, and the resonances were
    observed experimentally \cite{Ahmed:2012nh,Souza:2014sh}. Furthermore, in cosmology we encounter models
     in which our universe is result of continuous collision of branes and nucleation and
     therefore splitting of branes is a fundamental scenario in these models
     \cite{Khoury:2001wf}-\cite{Gen:2001bx}.\\

      Therefore, the universal aspects of thick brane splitting in warped bulk is
      important.
       such branes are constructed from a complex $\phi^{4}$ scalar
       field potential \cite{Campos:2001pr}, or from a real $\phi^{6}$ scalar field
       potential \cite{Zhao:2010mk}. These branes can be constructed
       from deformation of $\phi^{4}$ scalar field potential as well.\cite{Bazeia:2003qt}-\cite{Bazeia:1999}.\\
Recntly, Dutra and co-workers proposed a new model of thick brane in
which multi-brane scenario arises from scalar
 field models generating usual kink solutions \cite{Dutra:2014xla}. It suggests a special type of brane splitting. In this method superpotential
 function and warp factor will decompose in a special form and field equations will be simplified significantly.
 In this work we deal with this thick brane model
 which arises from polynomial superpotential.\\
The localization of spin $1/2$ fermions on the brane is very
interesting and important. Usually, in five dimension fermion do not
have a normalizable zero mode without scalar-fermion coupling
\cite{Bajc:1999mh}-\cite{Koley:2008dh}. In five dimension, with
 a Yukawa scalar-fermion coupling there may exist a massless bound state and a
 continuous
  gapless spectrum
 of massive Kaluza-Klein (KK) states  \cite{Liu:2007ku,BarbosaCendejas:2005kn}, while in some of other brane models,
  there exist some discrete KK state and a continuous gapless mass spectrum
  starting
   from positive $m^{2}$ \cite{BarbosaCendejas:2007vp,Liu:2008wd}.\\
This paper is organized as follows. In the next section, we present the brane model that
 is constructed in Ref. \cite{Dutra:2014xla} In section \ref{sec:zero-localize}.
 we investigate localization of the zero mode of the fermion field on the brane
 which is
 derived from polynomial potential. In section \ref{sec:massive-localize}
  we study localization of massive fermionic mode .
  Finally, in the last section we present our conclusions.\\

\section{The model}
\label{sec:model}

 we consider the following action in which  two
scalar fields coupled to gravity in 5 dimensions
\begin{equation}
\label{eq:action}
S=\int d^{4}xdr\sqrt{-g}[\frac{1}{4}R-\frac{1}{2}\partial_{M}\phi_{1}\partial^{M}\phi_{1}-\frac{1}{2}\partial_{M}\phi_{2}\partial^{M}\phi_{2}-V(\phi_{1},\phi_{2})]
\end{equation}
where $g=\det(g_{MN})$, $M,N=0,1,2,3,4$. The coordinates in the brane is represented by $x^{\mu}$ ($\mu={0,1,2,3}$) while the  coordinate in the bulk is shown by $x^{4}=r$. the line element is written as
\begin{equation}
\label{eq:metric}
ds^{2}=g_{MN}dx^{M}dx^{N}=e^{2A(r)}\eta_{\mu\nu}dx^{\mu}dx^{\nu}+dr^{2},
\end{equation}
where $\eta_{\mu\nu}$ is usual M
inkowski metric with $diag(-,+,+,+)$
and $e^{2A(r)}$ is called the warp factor. For this braneworld
scenario, the equations of motion is obtained as
\begin{eqnarray}
\label{eq:eom1}
\phi_{i}^{\prime\prime}+4A^{\prime}\phi_{i}^{\prime}&=&\dfrac{dV}{d\phi_{i}}\,.
\qquad
(i=1,2)
\\
\label{eq:eom2}
A^{\prime\prime}+\frac{2}{3}(\phi_{1}^{\prime 2}+\phi_{2}^{\prime 2})&=&0\,.
\\
\label{eq:eom1}
A^{\prime 2}+\frac{1}{3}V(\phi_{1},\phi_{2})&=&\frac{1}{6}(\phi_{1}^{\prime 2}+\phi_{2}^{\prime 2}).
\end{eqnarray}

the potential $V(\phi_{1},\phi_{2})$ can be written in terms of a superpotential $W(\phi_{1},\phi_{2})$ as
\begin{equation}
\label{eq:V}
V(\phi_{1},\phi_{2})=\frac{1}{2}\sum_{i=1}^{2}\big(\dfrac{\partial W}{\partial \phi_{i}}\big)^{2}-\frac{4}{3}W^{2}
\end{equation}
Therefore, equations of motion can be reduced to the following first order equations
\begin{equation}
\label{eq:eomlinear1}
\dfrac{d\phi_{i}}{dr}=\dfrac{\partial W}{\partial \phi_{i}},
\qquad
\dfrac{dA}{dr}=-\frac{2}{3}W.
\qquad
(i=1,2)
\end{equation}
But
\begin{equation}
\label{eq:W12}
W(\phi_{1},\phi_{2})=W_{1}(\phi_{1})+W_{2}(\phi_{2}),
\qquad
A(r)=A_{1}(r)+A_{2}(r)
\end{equation}
the first order equations in \eqref{eq:eomlinear1} are converted to
\begin{equation}
\label{eq:eomlinear2}
\dfrac{d\phi_{i}}{dr}=\dfrac{\partial W_{i}(\phi_{i})}{\partial \phi_{i}}\,
\qquad
\dfrac{dA_{i}}{dr}=-\frac{2}{3}W_{i}(\phi_{i}).
\qquad
(i=1,2)
\end{equation}
for polynomial superpotential
\begin{equation}
\label{eq:diffw}
W_{i}(\phi_{i})=\lambda_{i} (\phi_{i}-\dfrac{\phi_{i}^{3}}{3})
\qquad
(i=1,2)
\end{equation}
$\phi_{i}$ and $A{i}$ is given by:
\begin{equation}
\label{eq:phi}
\phi_{i}=\tanh[\lambda_{i}(r-r_{i})]
\qquad
(i=1,2)
\end{equation}
\begin{equation}
\label{eq:warp}
A(r)=\frac{1}{9}\sum_{i=1}^{2}
\{
\tanh^{2}[\lambda_{i}(\tilde{r}-r_{i})]-\tanh^{2}[\lambda_{i}(r-r_{i})
\}
-\frac{4}{9}\ln
(\prod_{i=1}^{2}\dfrac{sech[\lambda_{i}(\tilde{r}-r_{i})]}{sech[\lambda_{i}(r-r_{i})]}
\end{equation}
where $r_{i}$ is an integration constant, representing the center of
the kink and $\tilde{r}$ is defined as the average value of
coordintes of center of the kinks
\begin{equation}
\label{eq:r-average}
\tilde{r}=\frac{1}{2}(r_{1}+r_{2})
\end{equation}
In order to this brane model support brane splitting mechanism, we consider two symmetric kink solutions. Therefore, $\lambda_{1}=\lambda_{2}=\lambda$ and $r_{1}=-r_{2}=a$. So $\tilde{r}=0$ and we can write equations \eqref{eq:phi} and \eqref{eq:warp} as
\begin{equation}
\label{eq:phi12}
\phi_{1}=\tanh[\lambda(r-a)],
\qquad
\phi_{2}=\tanh[\lambda(r+a)]
\end{equation}
\begin{eqnarray}
\label{eq:A12}
A(r)&=&\frac{1}{9}
\{
2\tanh^{2}(\lambda a)-\tanh^{2}[\lambda (r-a)-\tanh^{2}[\lambda (r+a)
\} \nonumber \\
&-&\frac{4}{9}\ln
\left
(\dfrac{\cosh[\lambda(r-a)]\cosh[\lambda(r+a)]}{\cosh^{2}(\lambda a)}
\right)
\end{eqnarray}

\begin{figure}[htbp]
\centering
\includegraphics[scale=0.2]{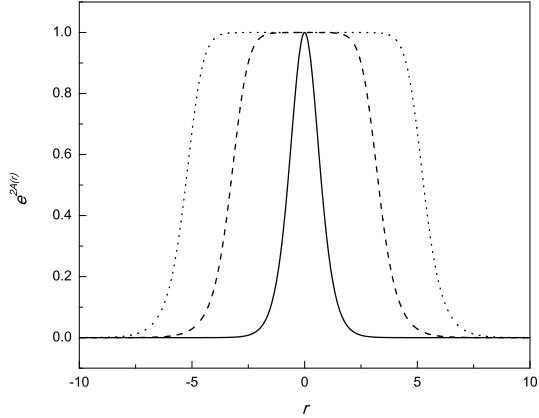}
\caption{ \label{fig:warp}\small plots of warp factor
$e^{2A(r)}$
for $a=0$ (solid line),
$a=3.0$ (dashed line) and
$a=5.0$ (dotted line). we put $\lambda=1$}
\end{figure}

Fig. \ref{fig:warp} represents warp factor for different values of $a$.
For $a=0$ we have a single brane that warp factor has a sharp peak.
For $a>0$ a plateau is formed in the very inside the brane where the energy density vanishes.
This is attributed to presence a new phase inside the brane. The plateau region inside the brane grows when $a$ increase.\\
In the next section, we investigate localization of zero mode of the fermion on the brane.
\section{Localization of zero mode}
\label{sec:zero-localize} Let us consider the action of fermion
coupled to gravity and the background scalar fields
\begin{equation}
\label{eq:ftionac}
S_{f}=\int d^{5}x \sqrt{-g} [\bar{\Psi}\Gamma^{M}D_{M}\Psi-\eta \bar{\Psi}F(\phi_{1},\phi_{2})\Psi],
\end{equation}
where $\eta$ is coupling constant. Here the back-
reaction effect of
fermion on background scalar field is neglected and scalar field is
considered to be unchanged. $\Gamma_M$ is general gamma matrics,
$\Gamma^{M}=(e^{-A}\gamma^{\mu},e^{-A}\gamma^{4})$.
\\ By following coordinate transformation
\begin{equation}
\label{eq:coordinatechange}
dz=e^{-A}dr
\end{equation}
the metric in \eqref{eq:metric} is changed to conformally flat one.
\begin{equation}
\label{eq:conformal-metric}
ds^{2}=e^{2A}(\eta_{\mu\nu}dx^{\mu}dx^{\nu}+dz^{2})
\end{equation}
 the equation of motion for fermion is derived as
\begin{equation}
\label{eq:feom}
[\gamma^{\mu}\partial_{\mu}+\gamma^{4}(\partial_{z}+2\partial_{z}A)-\eta e^{A}F(\phi_{1},\phi_{2})]\Psi=0.
\end{equation}
For solving this equation, we separate variables with KK and chiral decomposition
\begin{equation}
\label{eq:separVar}
\Psi(x,z)=\sum_{n}
\Big(\psi_{Ln}(x)f_{Ln}(z)+\psi_{Rn}(x)f_{Rn}(z)\Big)
\end{equation}
The 4D left-handed and right-handed fermions satisfy the Dirac
equations
\begin{equation}
\label{eq:Gamma}
\begin{split}
\gamma^{\mu}\partial_{\mu}\psi_{Ln}=m_{n}\psi_{Rn}
\\
\gamma^{\mu}\partial_{\mu}\psi_{Rn}=m_{n}\psi_{Ln}
\end{split}
\end{equation}
while the KK modes satisfy
\begin{equation}
\label{eq:KK1}
\{
\partial_{z}+2\partial_{z}A+\eta e^{A}F(\phi_{1},\phi_{2})
\}
f_{Ln}(z)=m_{n}f_{Rn}(z)
\end{equation}
\begin{equation}
\label{eq:KK2}
\{
\partial_{z}+2\partial_{z}A-\eta e^{A}F(\phi_{1},\phi_{2})
\}
f_{Rn}(z)=-m_{n}f_{Ln}(z)
\end{equation}
with the following ortho-normality condition
\begin{equation}
\label{eq:orthonormality}
\begin{split}
\int_{-\infty}^{\infty}dz e^{4A}f_{Ln}f_{Lm}=\int_{-\infty}^{\infty}dz e^{4A}f_{Rn}f_{Rm}=\delta_{nm},
\\
\int_{-\infty}^{\infty}dz e^{4A}f_{Ln}f_{Rm}=0
\end{split}
\end{equation}
and defining $\tilde{f}_{Ln}=e^{2A}f_{Ln}$ and $\tilde{f}_{Rn}=e^{2A}f_{Rn}$, the Schr\"{o}dinger-like equations are obtained.
\begin{equation}
\label{eq:ScheqL}
[-\partial_{z}^{2}+V_{L}(z)
]
\tilde{f}_{Ln}=m_{n}^{2}\tilde{f}_{Ln}
\end{equation}
\begin{equation}
\label{eq:ScheqR}
[-\partial_{z}^{2}+V_{R}(z)
]
\tilde{f}_{Rn}=m_{n}^{2}\tilde{f}_{Rn}
\end{equation}
where the effective potentials are given by
\begin{equation}
\label{eq:VSch-z}
\begin{split}
V_{L}(z)=\eta^{2}e^{2A}F^{2}(\phi_{1},\phi_{2})-\eta \partial_{z}[e^{A}F(\phi_{1},\phi_{2})]
\\
V_{R}(z)=\eta^{2}e^{2A}F^{2}(\phi_{1},\phi_{2})+\eta \partial_{z}[e^{A}F(\phi_{1},\phi_{2})]
\end{split}
\end{equation}
Because of the complexity of $A(r)$, we can not use relation \eqref{eq:coordinatechange} to obtain analytical form of $z(r)$. Therefore we apply numerical method  to get pair of $(r,z)$. with realations
\begin{equation}
\label{eq:AFz}
\partial_{z}A=e^{A(r)}\partial_{r}A,
\qquad
\partial_{z}F=e^{A(r)}\partial_{r}F
\end{equation}
By these relations, we can rewrite the potentials as the function of
$r$
\begin{equation}
\label{eq:VSch-r}
\begin{split}
V_{L}(z(r))=\eta e^{2A}[\eta F^{2}-\partial_{r}F-F\partial_{r}A(r)]
\\
V_{R}(z(r))=\eta e^{2A}[\eta F^{2}+\partial_{r}F+F\partial_{r}A(r)]
\end{split}
\end{equation}
With substituting $m_{n}=0$ in \eqref{eq:KK1} and \eqref{eq:KK2}, the left-handed and right-handed zero mode of fermion are obtained as:
\begin{eqnarray}
\label{mzL0}
    f_{L0}\propto \exp \left[-\eta\int_{0}^{z}dz^{\prime}\mathrm{e}^{A(z^{\prime})}F(\phi_{1},\phi_{2}) \right] \\
\label{mzR0}
    f_{R0}\propto \exp \left[\eta\int_{0}^{z}dz^{\prime}\mathrm{e}^{A(z^{\prime})}F(\phi_{1},\phi_{2}) \right]
\end{eqnarray}
The normalization condition for localizing zero mode of left-handed  fermions on the brane is
\begin{equation}
\label{eq:norm-conLz}
\int_{-\infty}^{\infty}dz\exp
\left(
-2\eta \int_{0}^{z}dz^{\prime}e^{-A(z^{\prime})}F(\phi_{1}(z^{\prime}),\phi_{2}(z^{\prime}))
\right)
< \infty
\end{equation}
in $r$ coordinate, we have:
\begin{equation}
\label{eq:norm-conLr}
\int_{-\infty}^{\infty}dr\exp
\left(
-A(r)-2\eta \int_{0}^{r}dr^{\prime}F(\phi_{1}(r^{\prime}),\phi_{2}(r^{\prime}))
\right)
< \infty
\end{equation}
For right handed this condition is achieved by replacing
$\eta\rightarrow -\eta$. For studying localization of zero mode of
fermion on the brane, we should determine the suitable form of
$F(\phi_{1},\phi_{2})$. In the following subsections we try to
detrmine this function using normalization condition.
\subsection{$F=\phi_{1}\phi_{2}$}
The integrand in \eqref{eq:norm-conLr} can be expressed as
\begin{eqnarray}
\label{eq:phi1phi2}
I_{0} &\equiv& \exp \Big(-A(r)-2\eta \int_{0}^{r}dr^{\prime}F(\phi_{1}(r^{\prime}),\phi_{2}(r^{\prime})) \Big)\nonumber \\
&=&\exp[-\frac{1}{9}
\{
2\tanh^{2}(\lambda a)-\tanh^{2}[\lambda (r-a)-\tanh^{2}[\lambda (r+a)
\} -2\eta r]\nonumber \\
&\times&
\left(\dfrac{\cosh[\lambda(r-a)]\cosh[\lambda(r+a)]}{\cosh^{2}(\lambda a)}
\right)^{\frac{4}{9}}
\left
(\dfrac{1+e^{-2\lambda(x-a)}}{1+e^{-2\lambda(x+a)}}
\right)^{\frac{-2\eta}{\lambda}}\nonumber \\
\end{eqnarray}
We can decompose normalization condition to two regions as
\begin{equation}
\label{eq:I0-tworegion}
\int_{0}^{\infty} dr I_{0} < 0, \qquad \int_{-\infty}^{0} dr I_{0} < 0
\end{equation}
from first integral we have
\begin{equation}
\label{eq:I0-firstint}
I_{0}\rightarrow \exp(8/9-2\eta r) \qquad when \qquad r\rightarrow +\infty
\end{equation}
For satisfying normalization condition,  we require $\eta >4/9$. For second integral we have
\begin{equation}
\label{eq:I0-secondint}
I_{0}\rightarrow \exp(-8/9-2\eta r)\rightarrow \infty \qquad when \qquad r\rightarrow -\infty
\end{equation}
therefore the second integral is divergent. So the left-handed
fermion zero mode can not be localized on the brane. In the other
hand, with $\eta \rightarrow -\eta$, we can see that the zero mode
of right handed fermion can not localize too.
\subsection{$F=\phi_{1}-\phi_{2}$}
the integrand in \eqref{eq:norm-conLr} is written as
\begin{eqnarray}
\label{eq:I1}
I_{1} &\equiv& \exp[-\frac{1}{9}
\{
2\tanh^{2}(\lambda a)-\tanh^{2}[\lambda (r-a)-\tanh^{2}[\lambda (r+a)
\} \nonumber \\
&\times&
\left(\dfrac{\cosh[\lambda(r-a)]cosh[\lambda(r+a)]}{\cosh^{2}(\lambda a)}
\right)^{\frac{4}{9}}
\left
(\dfrac{\cosh[\lambda(r-a)]}{\cosh[\lambda(r+a)]}
\right)^{-\frac{2\eta}{\lambda}}
\end{eqnarray}
from which we have
\begin{equation}
\label{eq:I1asym}
I_{1} \rightarrow \exp(\pm8r/9) \rightarrow \infty \qquad when \qquad r \rightarrow \pm\infty
\end{equation}
Hence, left-handed zero mode can not be localized on the brane in
this case. Moreover, because  $I_{1}$ is independent of $\eta$
therefore, we can conclude that right handed fermion can not be
localized on the brane.
\subsection{$F=\phi_{1}+\phi_{2}$}
\begin{figure}[htb]
\begin{center}
\includegraphics[scale=0.2]{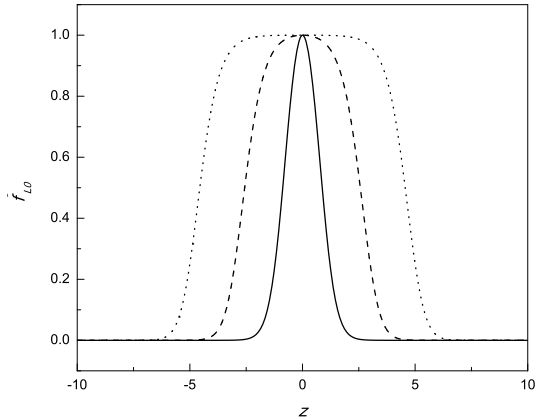}
\end{center}
\caption{\label{fig:zeromodes}Fermion zero modes localized on the brane : $a=1$ (solid line), $a=3$ (dashed line) and $a=5$
(dotted line).}
\end{figure}

The integrand in \eqref{eq:norm-conLr} is expressed as
\begin{eqnarray}
\label{eq:I2}
I_{2} &\equiv& \exp[-\frac{1}{9}
\{
2\tanh^{2}(\lambda a)-\tanh^{2}[\lambda (r-a)-\tanh^{2}[\lambda (r+a)
\} \nonumber \\
&\times&
\left(\dfrac{\cosh[\lambda(r-a)]\cosh[\lambda(r+a)]}{\cosh^{2}(\lambda a)}
\right)^{\frac{4}{9}-\frac{2\eta}{\lambda}}
\end{eqnarray}
therefore, we have:
\begin{equation}
\label{eq:I2asym}
I_{2} \rightarrow \exp[\pm(8r/9-4\eta r/\lambda)]  \qquad when \qquad r \rightarrow \pm \infty
\end{equation}
So the normalization condition for localization of left handed
fermion zero mode is
\begin{equation}
\label{eq:normconsum-L}
\eta > \frac{2\lambda}{9}
\end{equation}
By changing $\eta \rightarrow -\eta$ one can find that the zero mode
of right handed fermion, however this mode can not be localized on
the brane.
\\
The effective potentials for left handed and right handed KK fermion
have the form
\begin{eqnarray}
\label{eq:VLsum}
V_{L} &=& \eta \exp
\left[
\frac{2}{9}
\{
2\tanh^{2}(\lambda a)-\tanh^{2}[\lambda (r-a)]-\tanh^{2}[\lambda (r+a)]
\}
\right] \nonumber \\
&\times&
\left(
\dfrac{\cosh\lambda(r-a)]\cosh[\lambda(r+a)]}{\cosh^{2}(\lambda a)}
\right)^{-\frac{8}{9}}
\Big\{
\eta \big(\tanh[\lambda (r-a)]+\tanh[\lambda (r+a)] \big)^{2}\nonumber \\
&-&\lambda
\Big(
\sec \!h^{2}[\lambda (r-a)]+\sec\!h^{2}[\lambda (r+a)]+\frac{2}{3}\big(\tanh[\lambda (r-a)]+\tanh[\lambda (r+a)]\big) \nonumber \\
&\times&
\{
\tanh[\lambda (r-a)]+\tanh[\lambda (r+a)] \nonumber \\
&-&\dfrac{\tanh^{3}[\lambda (r-a)]+\tanh^{3}[\lambda (r+a)]}{3}
\}
\Big)
\Big\} \\
\label{eq:VRsum}
V_{R}&=&V_{L}|_{\eta\rightarrow-\eta}
\end{eqnarray}
The Values of potential in $r$ or $z=0$ and $r$ or
$z\rightarrow\pm\infty$ are
\begin{eqnarray}
\label{eq:VLsumasym}
&&V_{L}(0) = -V_{R}(0) =-2\eta \lambda \sec\!h^{2}(\lambda a) \\
\label{eq:VRsumasym}
&&V_{L}(\pm \infty) = V_{R}(\pm \infty) =0
\end{eqnarray}
\begin{figure*}[htb]
\begin{center}
\subfigure[]{\label{fig:leftpotential}
\includegraphics[width=5.5cm]{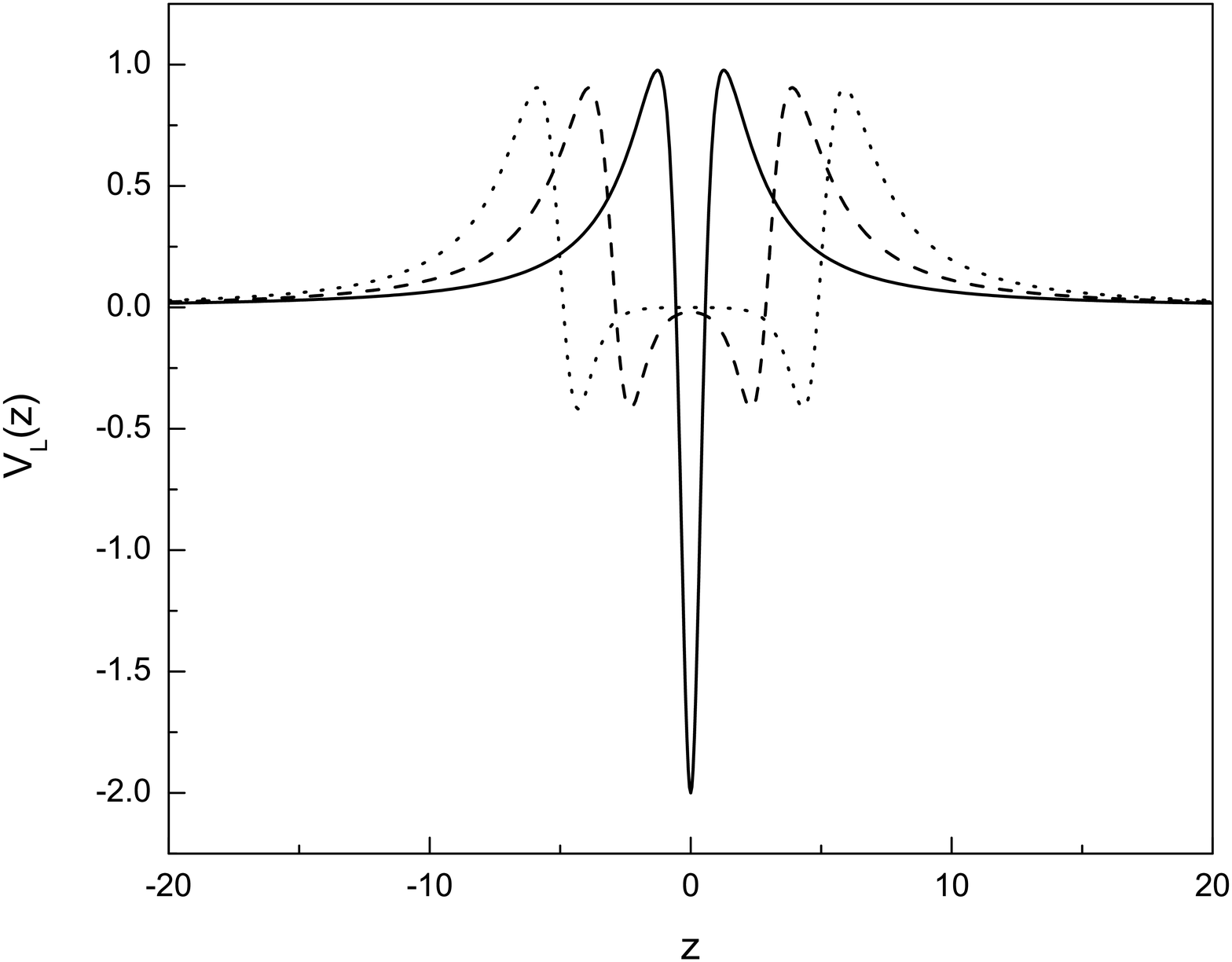}}
\subfigure[]{\label{fig:rightpotential}
\includegraphics[width=5.5cm]{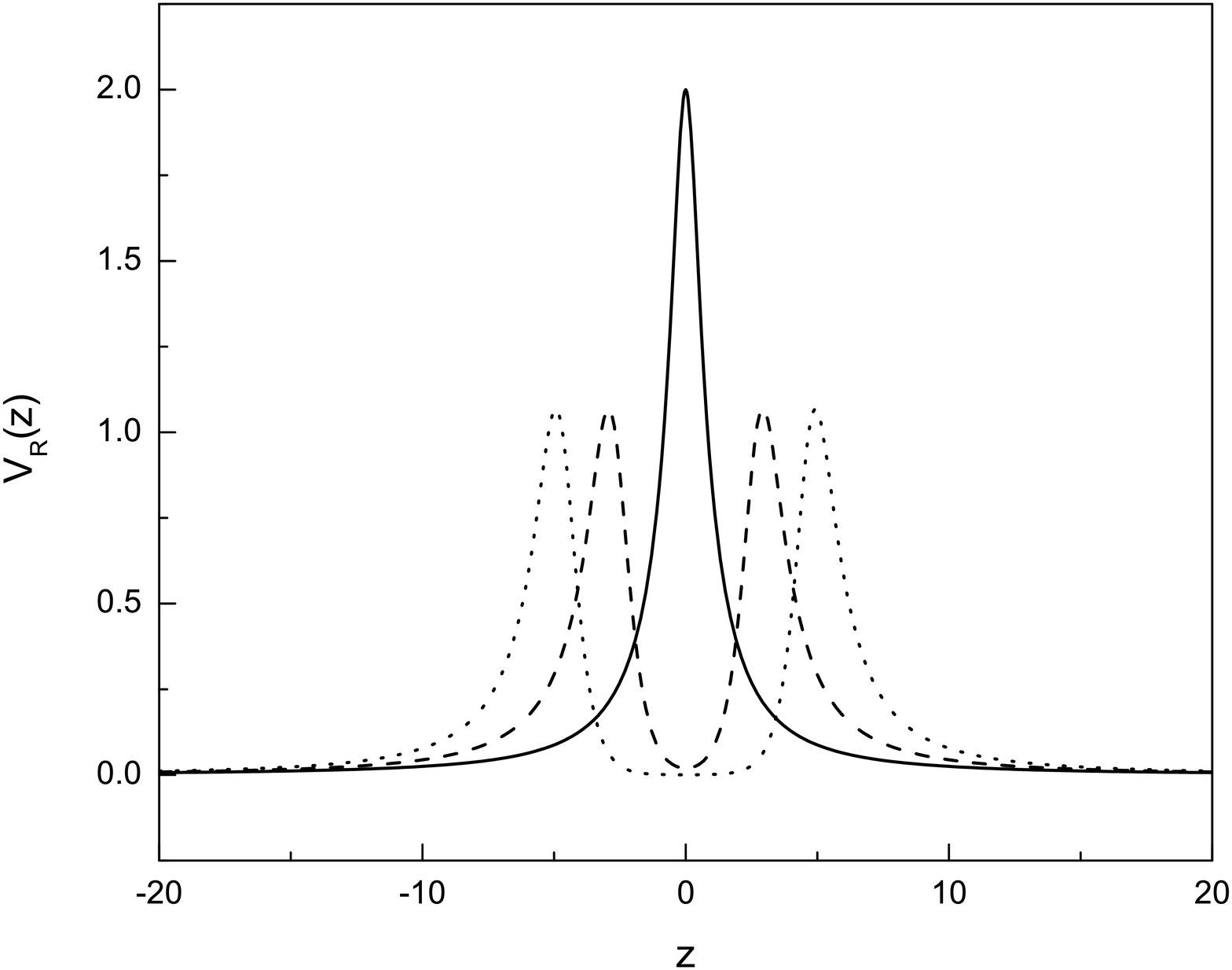}}
\end{center}
\caption{The shapes of potentials: (a) $V_{L}(z)$ and (b)
$V_{R}(z)$, $a=0$ (solid line), $a=3$ (dashed line) and $a=5$
(dotted line).}
\label{fig:potentials}
\end{figure*}
It can be seen that the asymptotic behaviors of two potentials are
the same when $y\rightarrow\pm\infty$, but opposite at the origin,
$z=0$. This reveals that only one of the massless left and right
chiral fermions can be localized on the brane. The shape of
effective potentials are shown in Fig. \ref{fig:potentials}. The
form of $V_{L}(z)$ is volcano type and therefore there is no mass
gap between the zero mode and KK excitation modes. On the other
hand, the $V_{R}(z)$ is always positive at the brane location. We
know that this type of potential can not trap any bound state of
right handed fermion and there is no zero mode of right handed
fermion.This is consistent with the pervious our knowledge that only
one chirality of massless fermion can exist.
\\
The zero mode of left-handed fermion is written as
\begin{eqnarray}
\label{eq:zeroLsum}
\tilde{f}_{L0}(z) &\varpropto& \exp\Big(-\eta \int_{0}^{z}dz^{\prime}e^{A(z^{\prime})}\big[\phi_{1}(z^{\prime})+\phi_{2}(z^{\prime})\big]\Big) \nonumber \\
&=& \exp\Big(-\eta \int_{0}^{r}dr^{\prime}\big[\phi_{1}(r^{\prime})+\phi_{2}(r^{\prime})\big]\Big) \nonumber \\
&=& \Big(
\cosh[\lambda(r-a)]\cosh[\lambda(r+a)]
\Big)^{-\frac{2\eta}{\lambda}}
\end{eqnarray}
Fig. \ref{fig:zeromodes} shows the form  of fermion zero modes on the brane. One can see that the width of the function is increased with splitting and the height is decreasd.
\subsection{The case $F=\phi_{1}+\beta \phi_{2}$}
The integrand in \eqref{eq:norm-conLr} is expressed as:
\begin{eqnarray}
\label{eq40}
I_{3} &\equiv& \exp[-\frac{1}{9}
\{
2\tanh^{2}(\lambda a)-\tanh^{2}[\lambda (r-a)-\tanh^{2}[\lambda (r+a)
\} \nonumber \\
&\times&
\left(\dfrac{\cosh[\lambda(r-a)]}{\cosh(\lambda a)}
\right)^{\frac{4}{9}-\frac{2\eta}{\lambda}}
\left(\dfrac{\cosh[\lambda(r+a)]}{\cosh(\lambda a)}
\right)^{\frac{4}{9}-\beta \frac{2\eta}{\lambda}}
\end{eqnarray}
Therefore, we have:
\begin{equation}
\label{eq:I3}
I_{3} \rightarrow \exp \big[\pm \big(8r/9-2(\beta +1)\eta r/\lambda \big)\big]  \qquad when \qquad r \rightarrow \pm\infty
\end{equation}
So the normalization condition becomes :
\begin{equation}
\label{eq:conGen}
\eta > \frac{4 \lambda}{9 (\beta +1)}
\end{equation}
For $\beta=-1$ the condition \eqref{eq:conGen} can not be satisfied,
Hence, left-handed zero mode can not be localized on the brane. for
$\beta=0$ we have $\eta>4\lambda/9$. This means that coupling of
fermion to every sub-brane can localize zero mode on the brane. For
the case $\beta=1$ the normalization condition is reduced to eq.
\eqref{eq:normconsum-L}. The zero mode of left-handed fermion is
turned out to be
\begin{eqnarray}
\label{eq:fLGen}
\tilde{f}_{L0}(z) &\varpropto& \exp\Big(-\eta \int_{0}^{r}dr^{\prime}[\phi_{1}(r^{\prime})+\beta \phi_{2}(r^{\prime})\Big) \nonumber \\
&=&
\cosh[\lambda(r-a)]^{-\frac{2\eta}{\lambda}}\cosh[\lambda(r+a)]^{-\frac{2\beta \eta}{\lambda}}
\end{eqnarray}
The explicit forms of the potentials are
\begin{eqnarray}
\label{eq:VGen}
V_{L} &=& \eta \exp
\left[
\frac{2}{9}
\{
2\tanh^{2}(\lambda a)-\tanh^{2}[\lambda (r-a)]-\tanh^{2}[\lambda (r+a)]
\}
\right] \nonumber \\
&\times&
\left(
\dfrac{\cosh\lambda(r-a)]\cosh[\lambda(r+a)]}{\cosh^{2}(\lambda a)}
\right)^{-\frac{8}{9}}
\Big\{
\eta \big(\tanh[\lambda (r-a)]+\beta \tanh[\lambda (r+a)] \big)^{2}\nonumber \\
&-&\lambda
\Big(
\sec \!h^{2}[\lambda (r-a)]+\beta^{2} \sec\!h^{2}[\lambda (r+a)]+\frac{2}{3}\big(\tanh[\lambda (r-a)]+\beta \tanh[\lambda (r+a)]\big) \nonumber \\
&\times&
\{
\tanh[\lambda (r-a)]+\tanh[\lambda (r+a)]-\dfrac{\tanh^{3}[\lambda (r-a)]+\tanh^{3}[\lambda (r+a)]}{3}
\}
\Big)
\Big\} \\
V_{R}&=&V_{L}|_{\eta\rightarrow-\eta}
\end{eqnarray}
\begin{figure*}[htbp]
\begin{center}
\includegraphics[width=5cm]{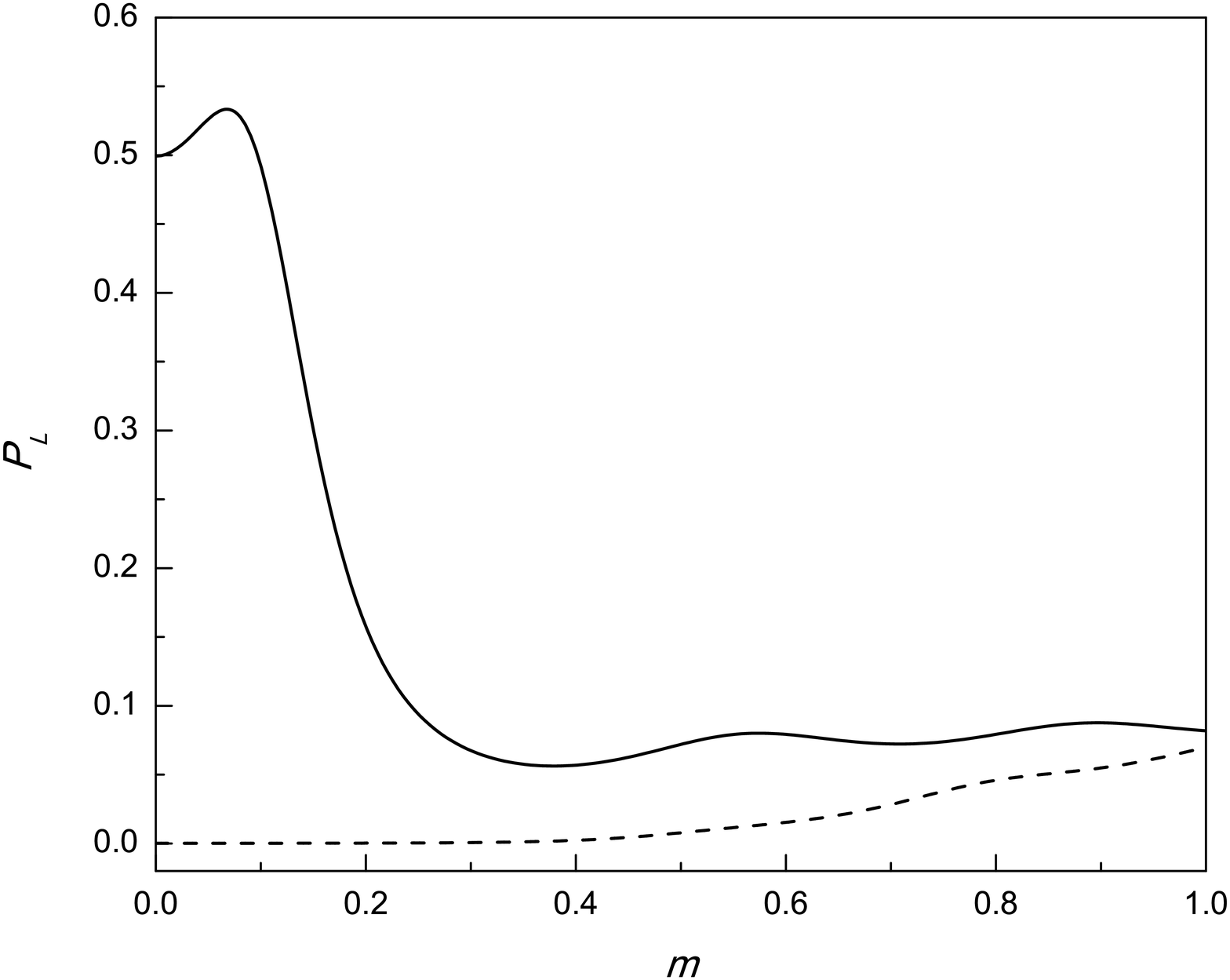}
\includegraphics[width=5cm]{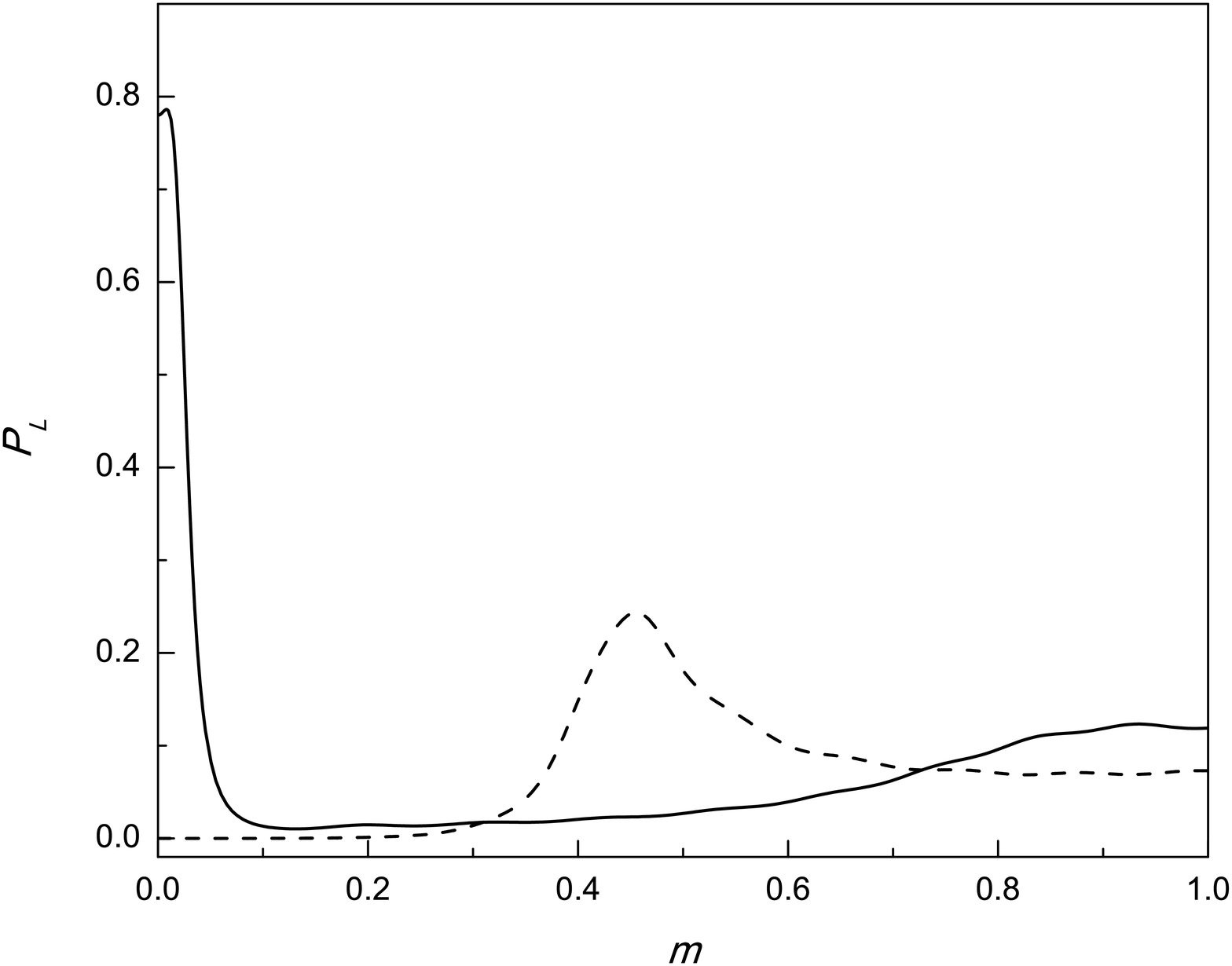}
\includegraphics[width=5cm]{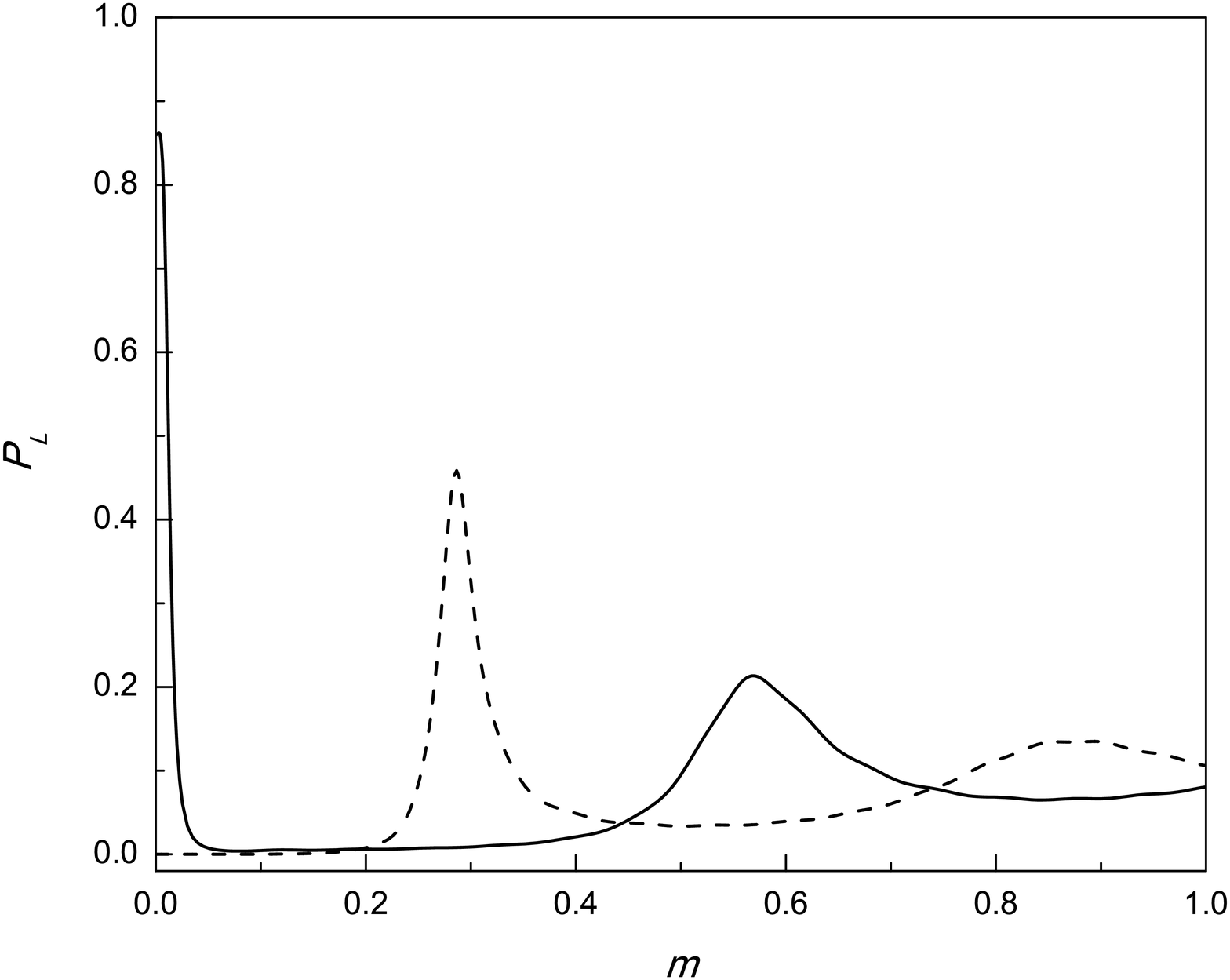}
\includegraphics[width=5cm]{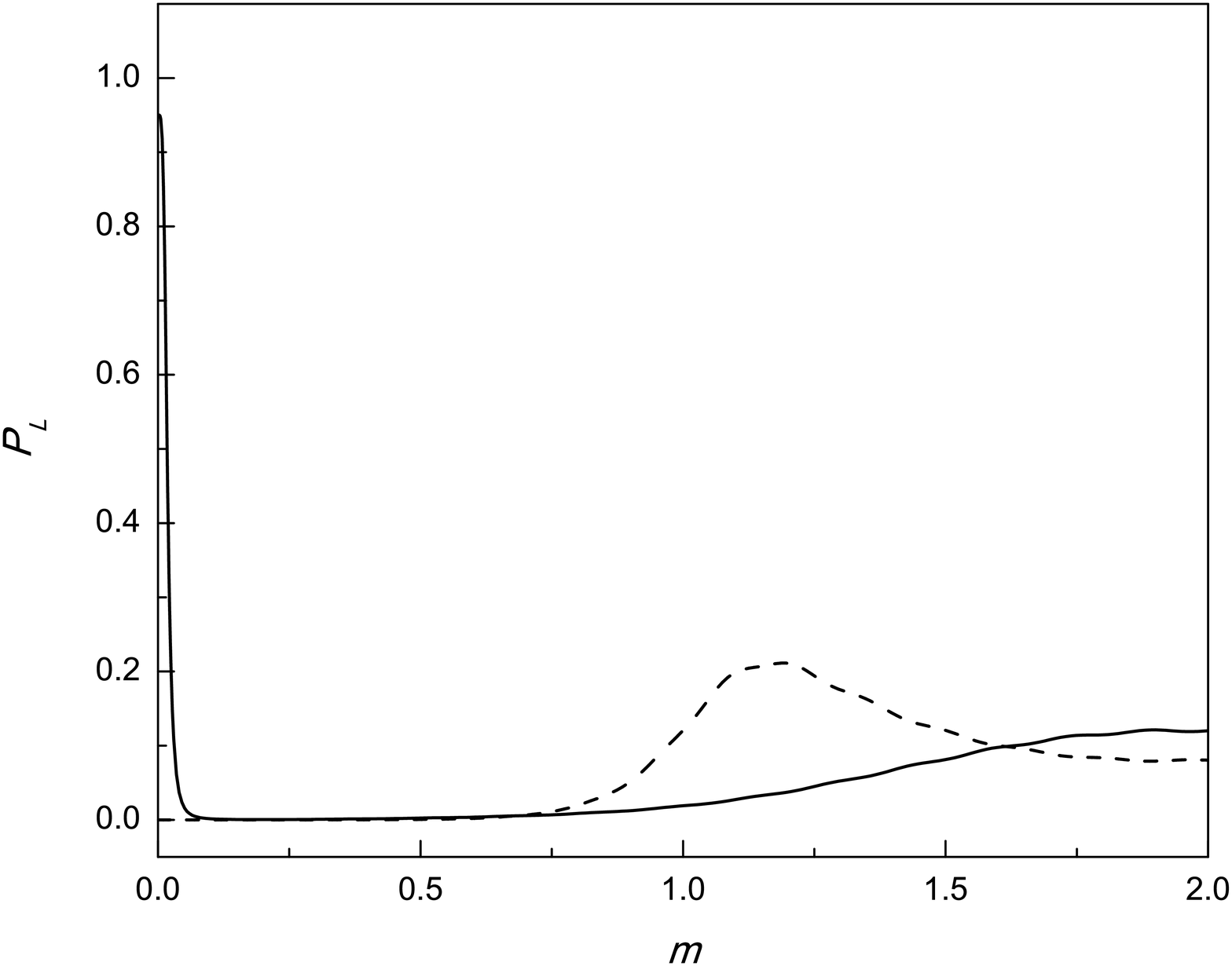}
\includegraphics[width=5cm]{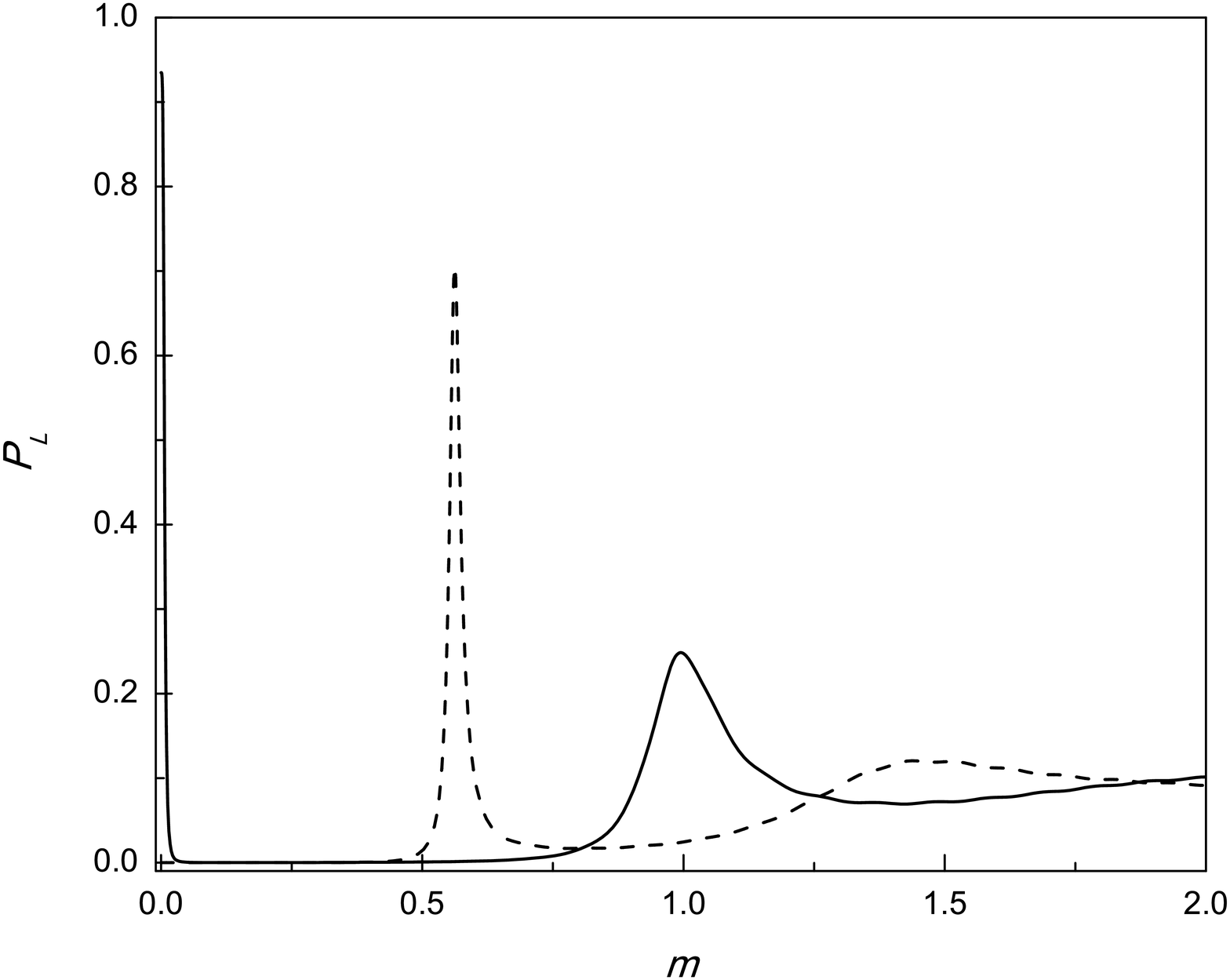}
\includegraphics[width=5cm]{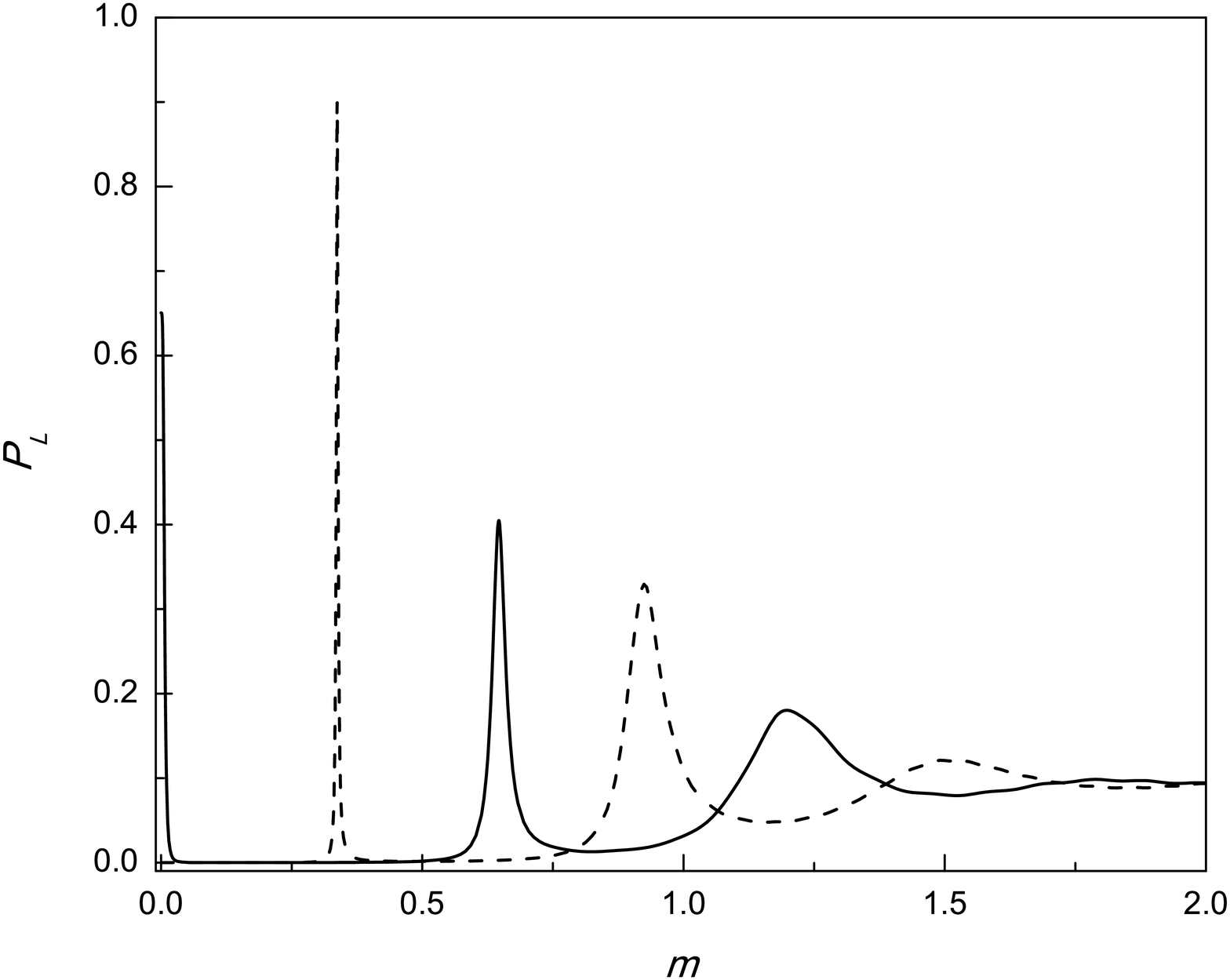}
\includegraphics[width=5cm]{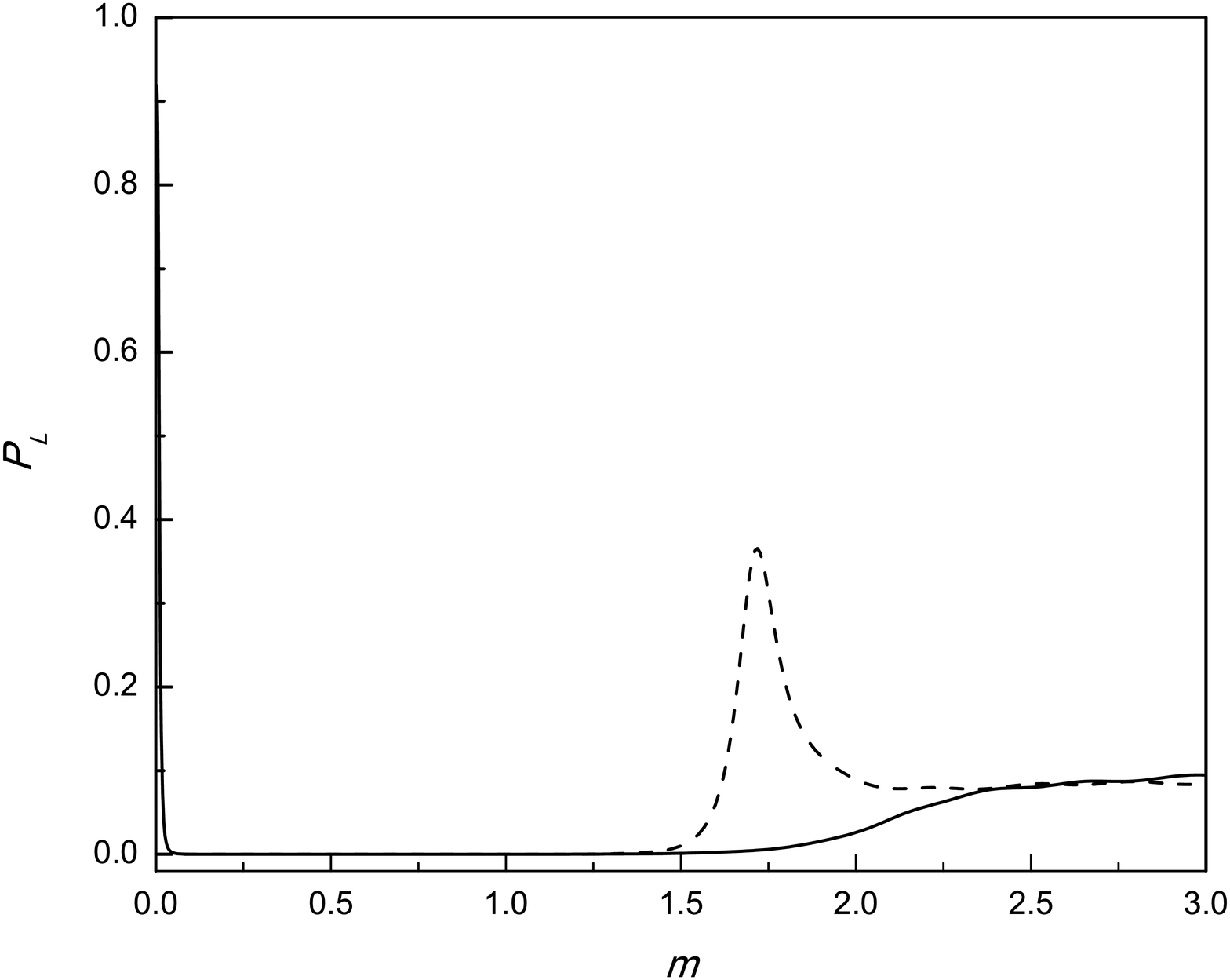}
\includegraphics[width=5cm]{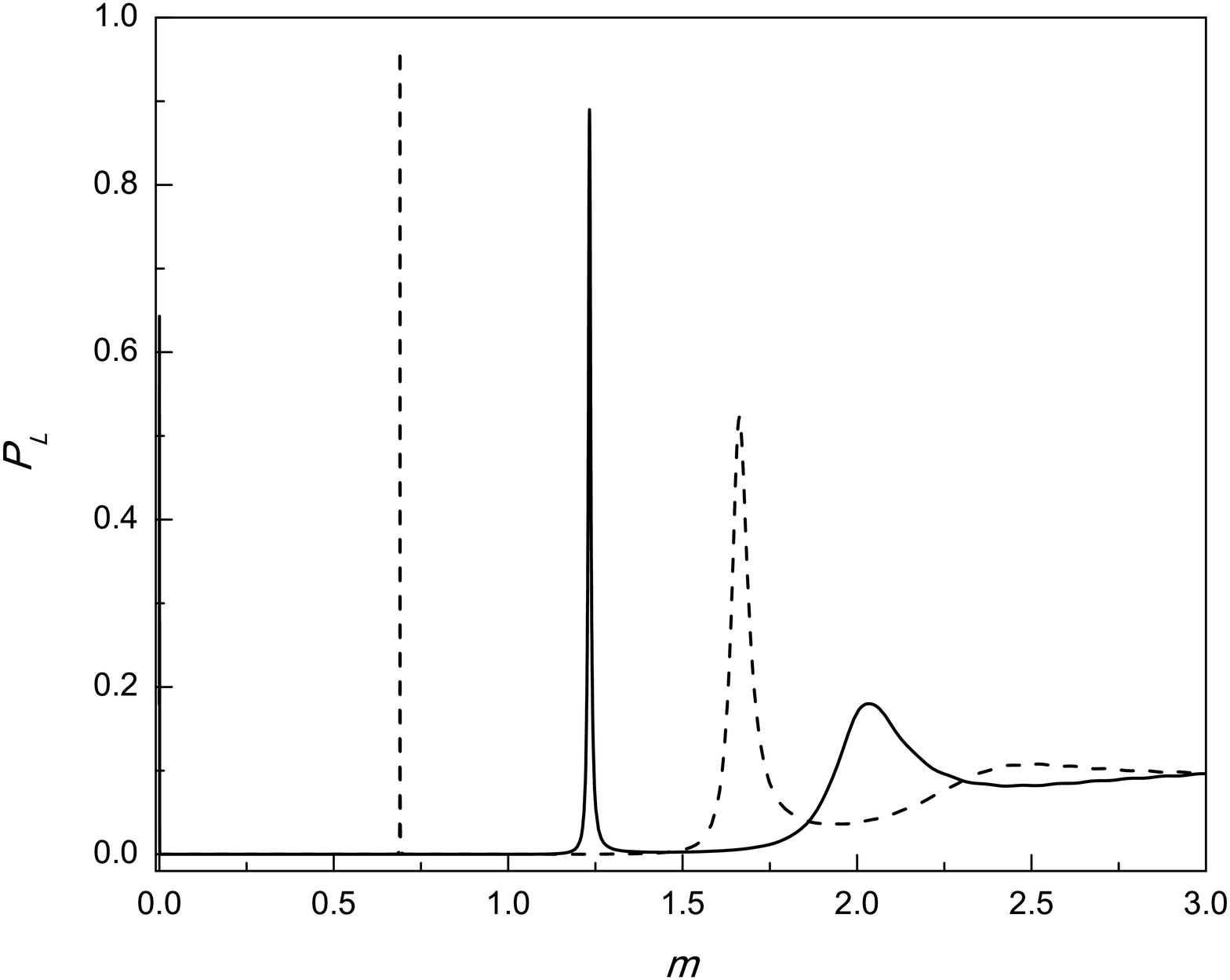}
\includegraphics[width=5cm]{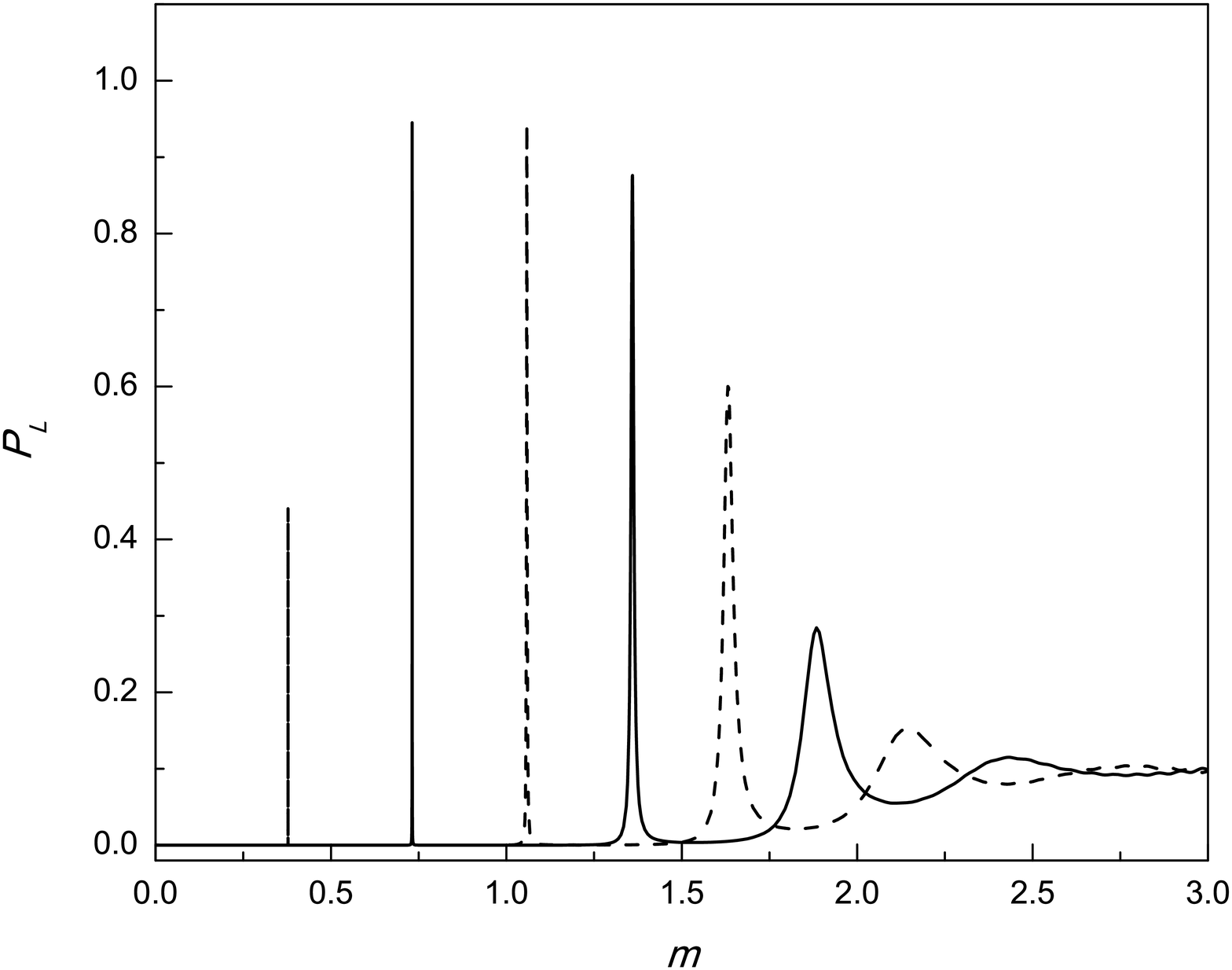}
\end{center}
\caption{\label{fig:mass-leftmodes} Plots of relative probability
for left
 handed fermion resonances with $a=1.0$(left column), $a=3.0$(middle column)
 and $a=5.0$(right column). Coupling constants are $\eta=0.5$(first raw),
 $\eta=1.0$(second raw) and $\eta=2.0$(third raw). Solid lines and dashed lines corresponds to even and odd parity respectively. In all cases $\lambda=1.0$.}
\end{figure*}
\begin{figure*}[htb]
\begin{center}
\includegraphics[width=5cm]{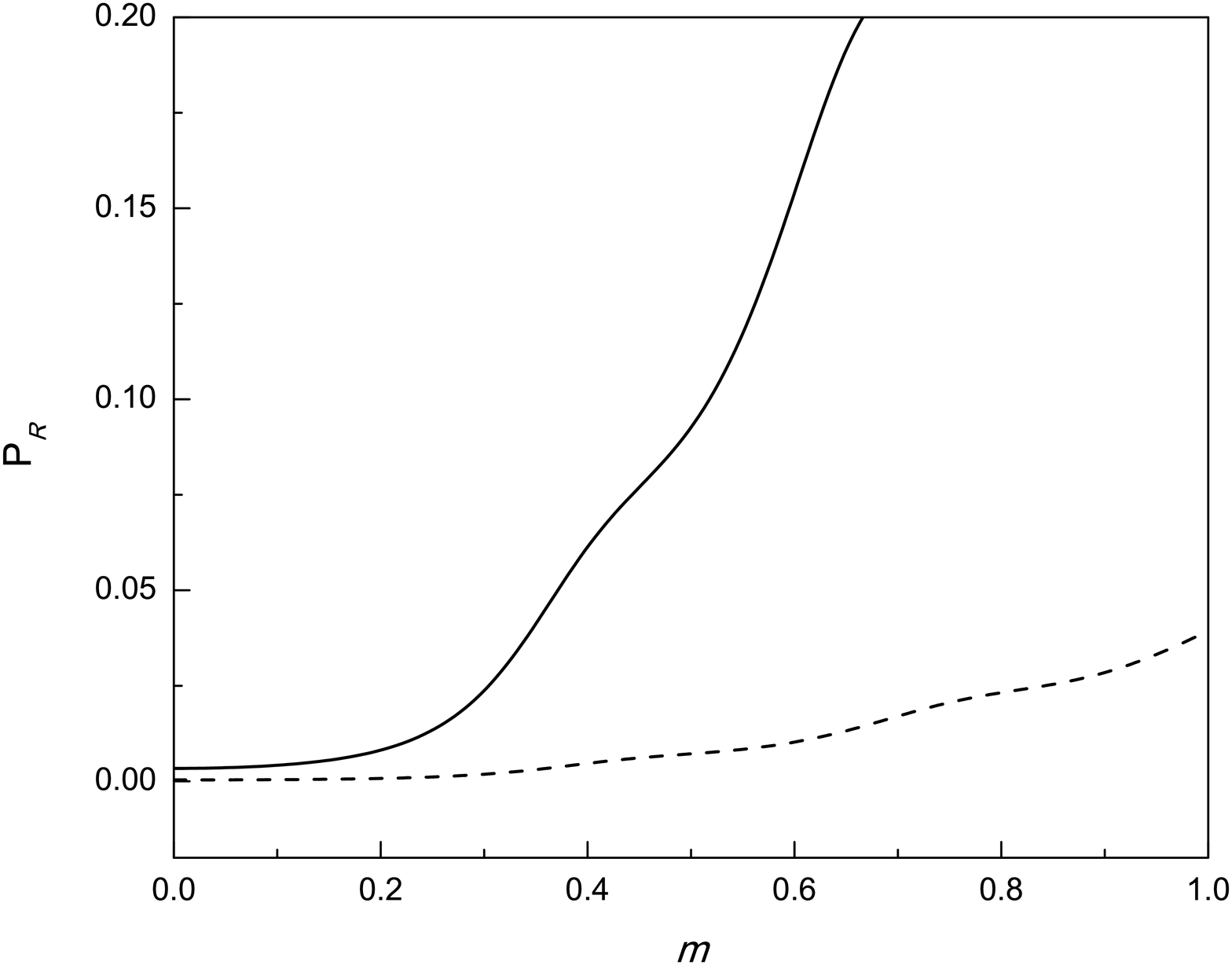}
\includegraphics[width=5cm]{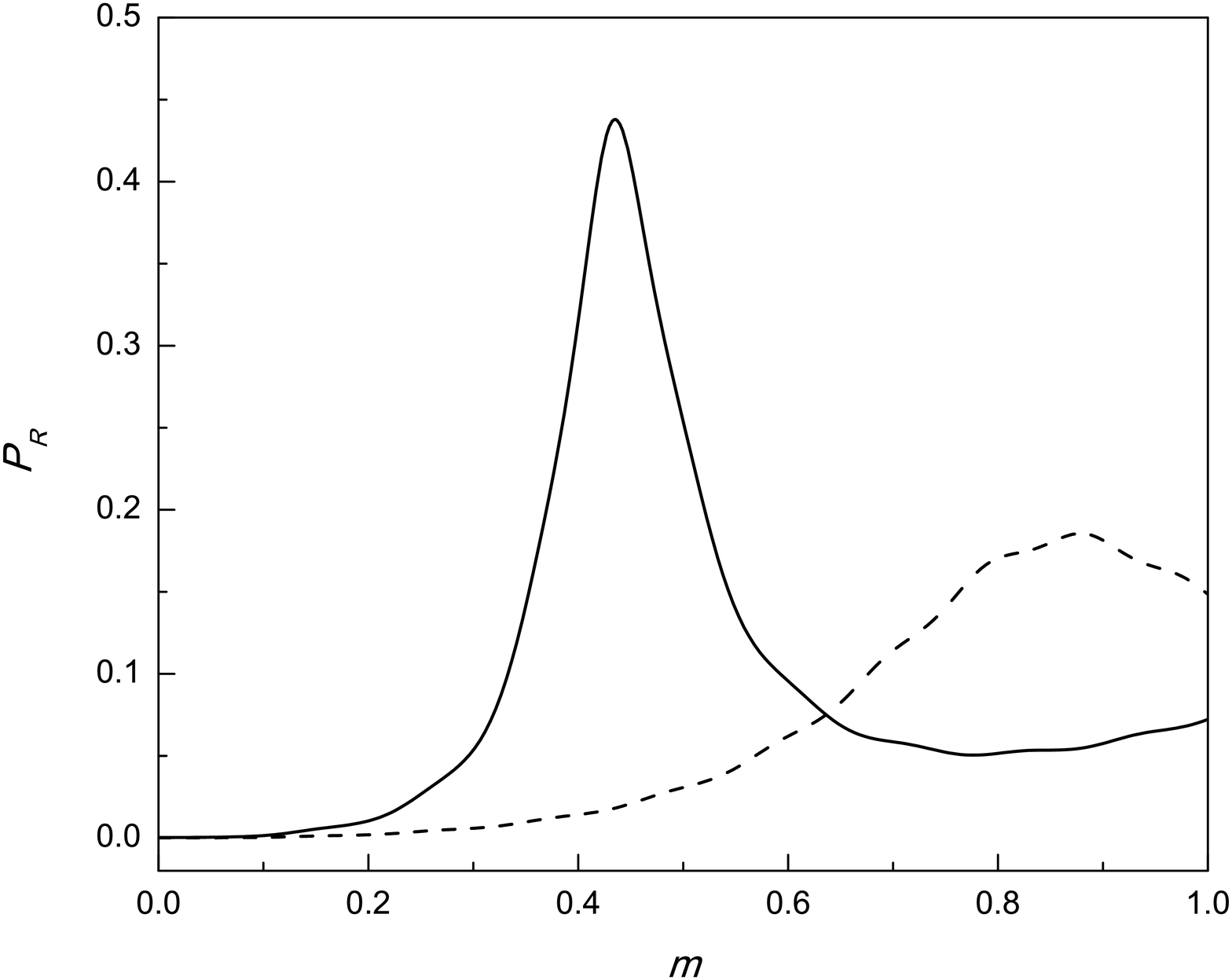}
\includegraphics[width=5cm]{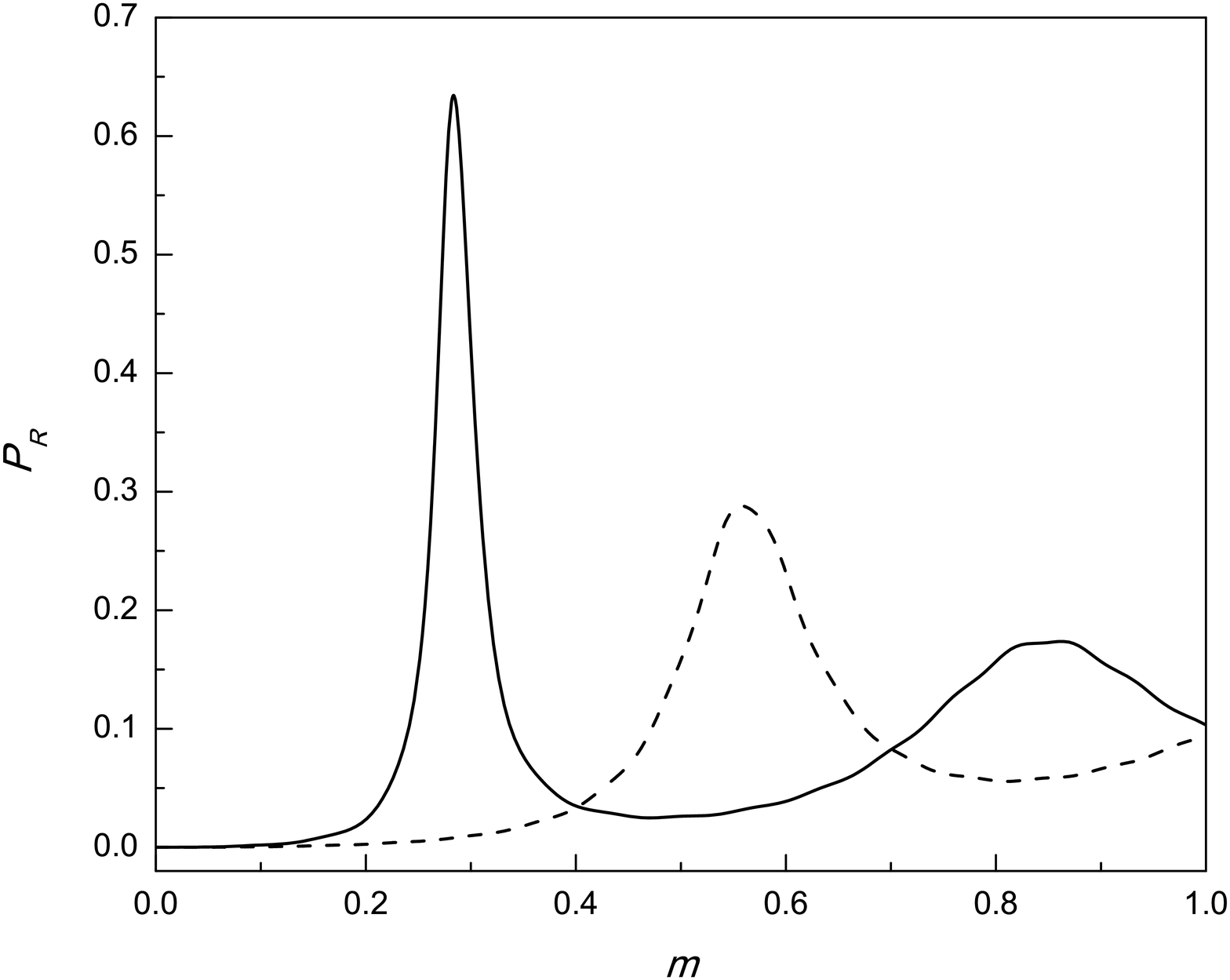}
\includegraphics[width=5cm]{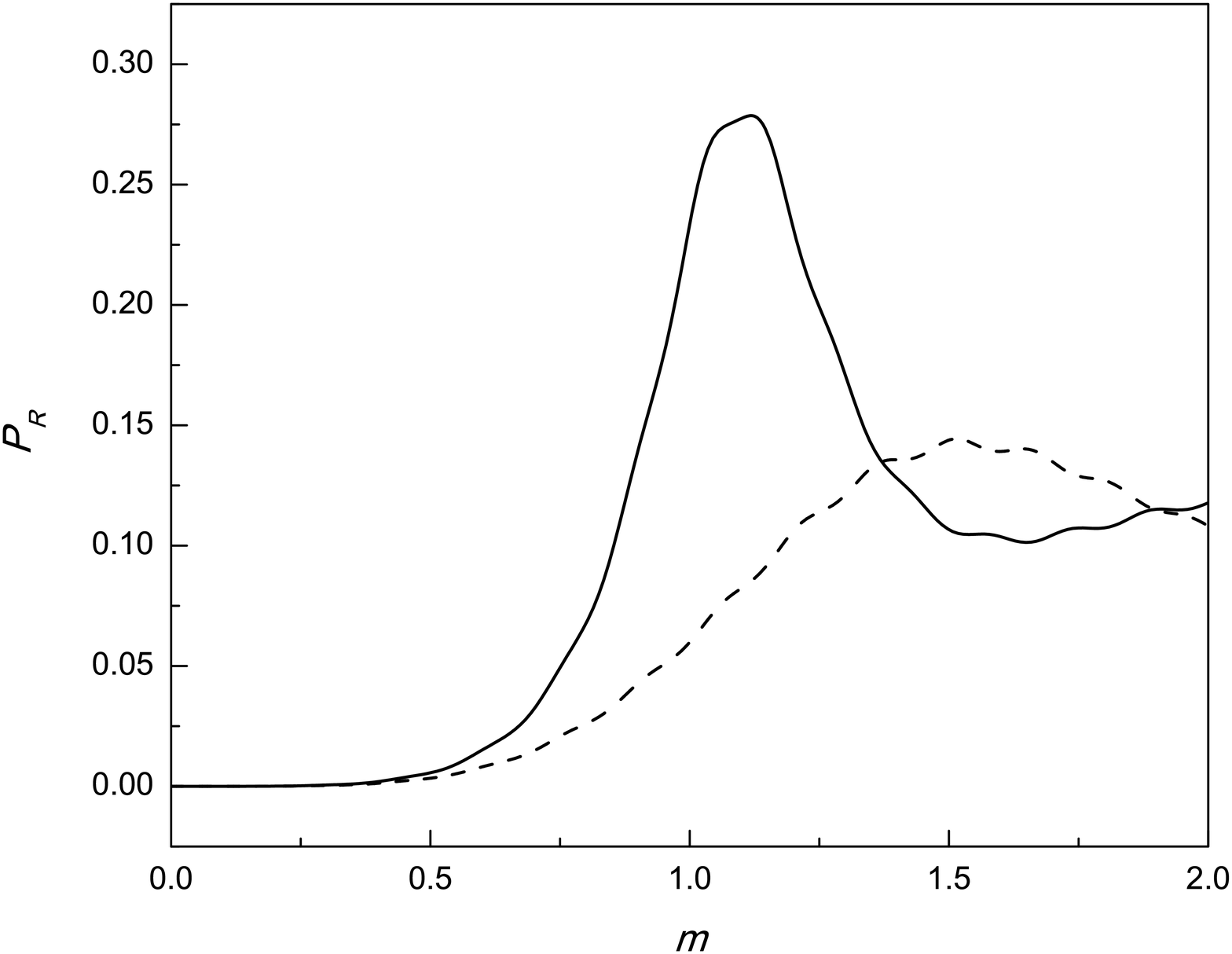}
\includegraphics[width=5cm]{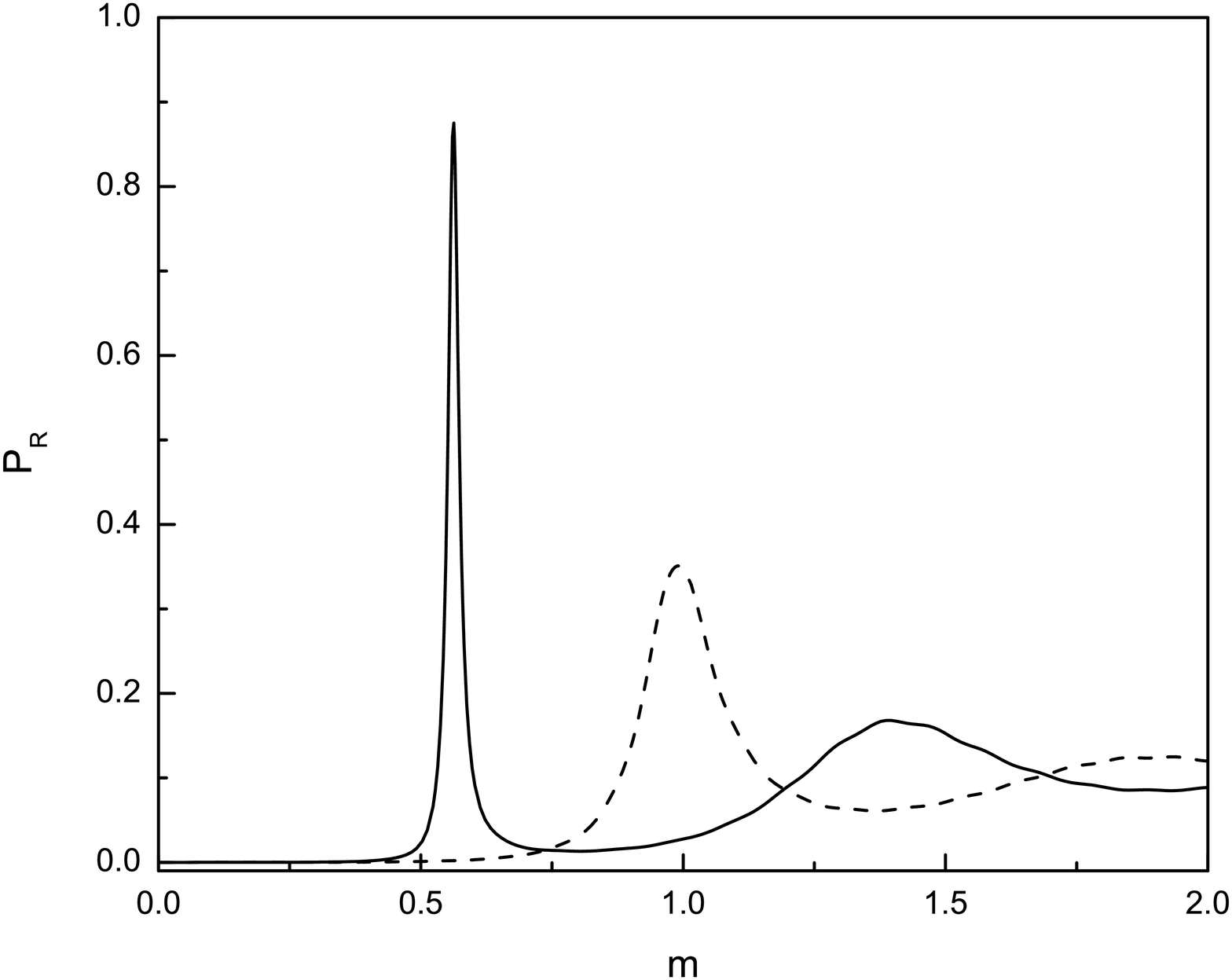}
\includegraphics[width=5cm]{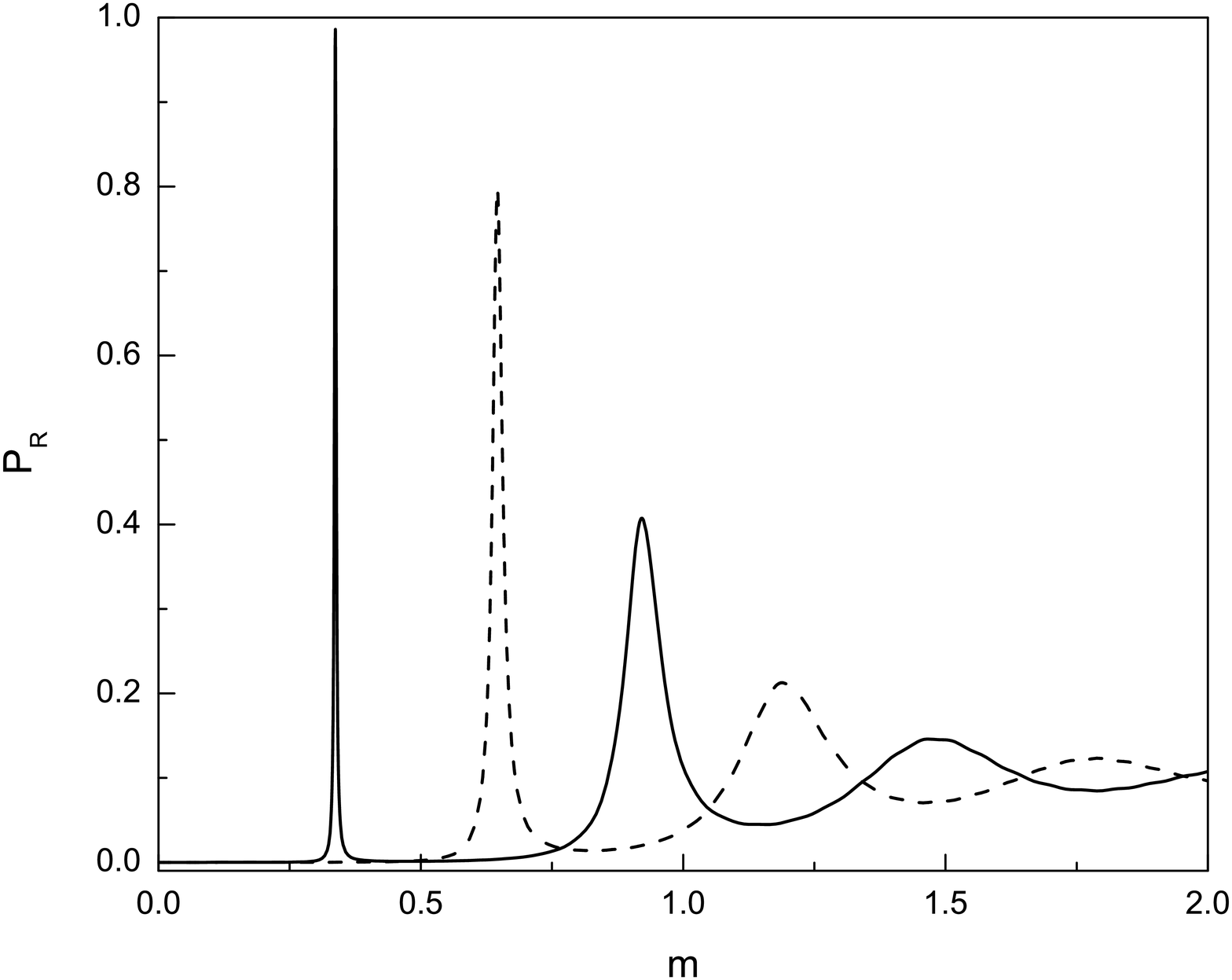}
\includegraphics[width=5cm]{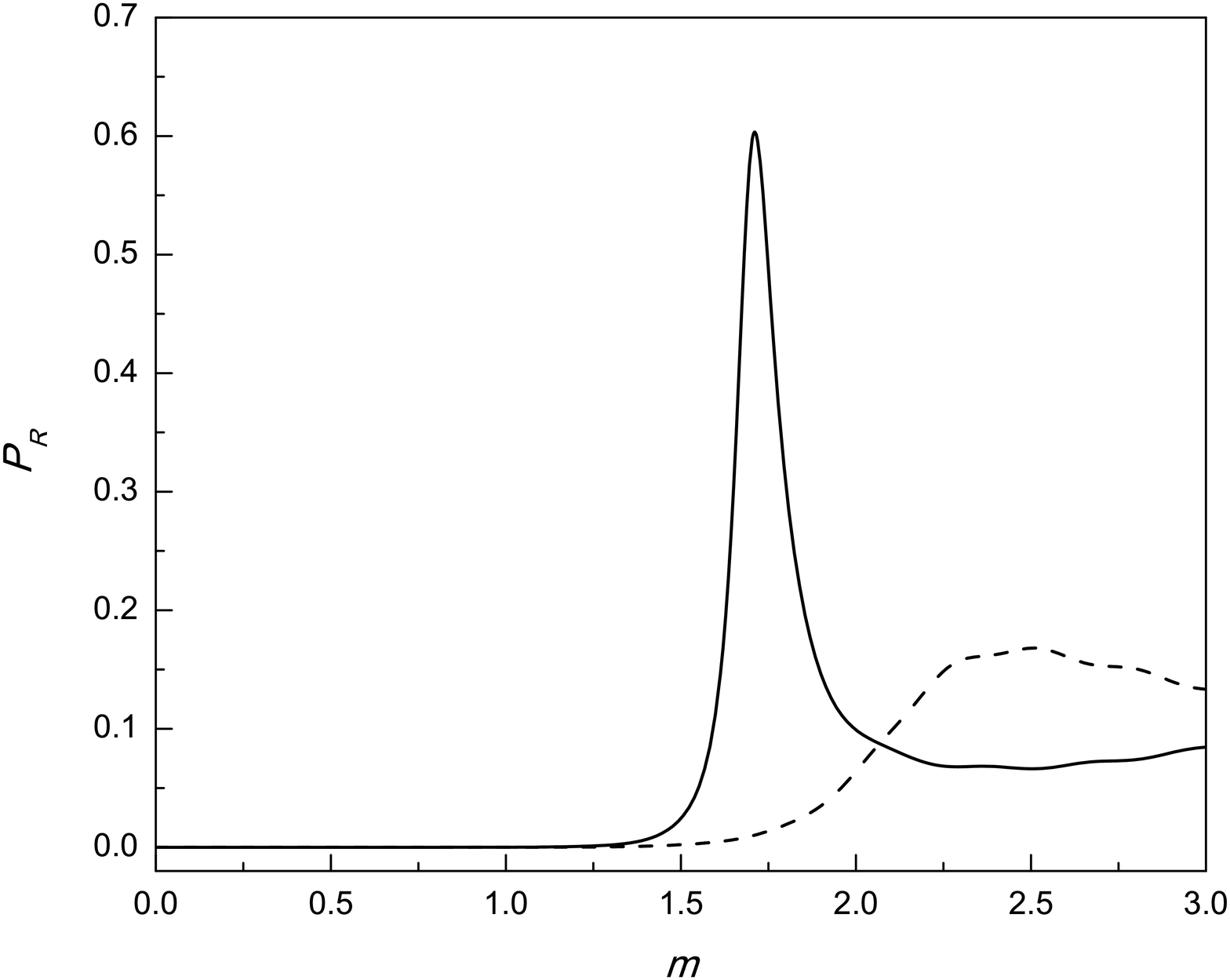}
\includegraphics[width=5cm]{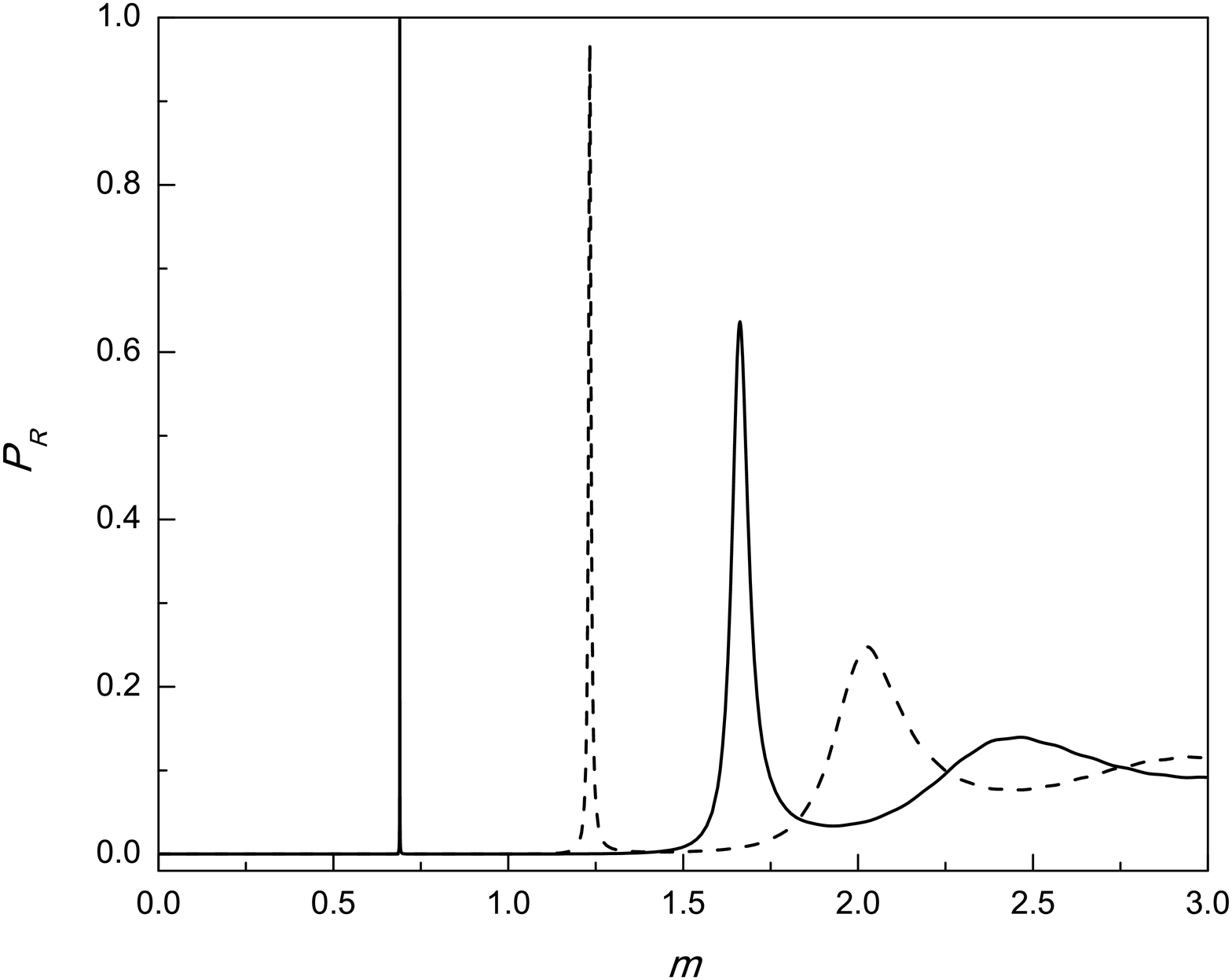}
\includegraphics[width=5cm]{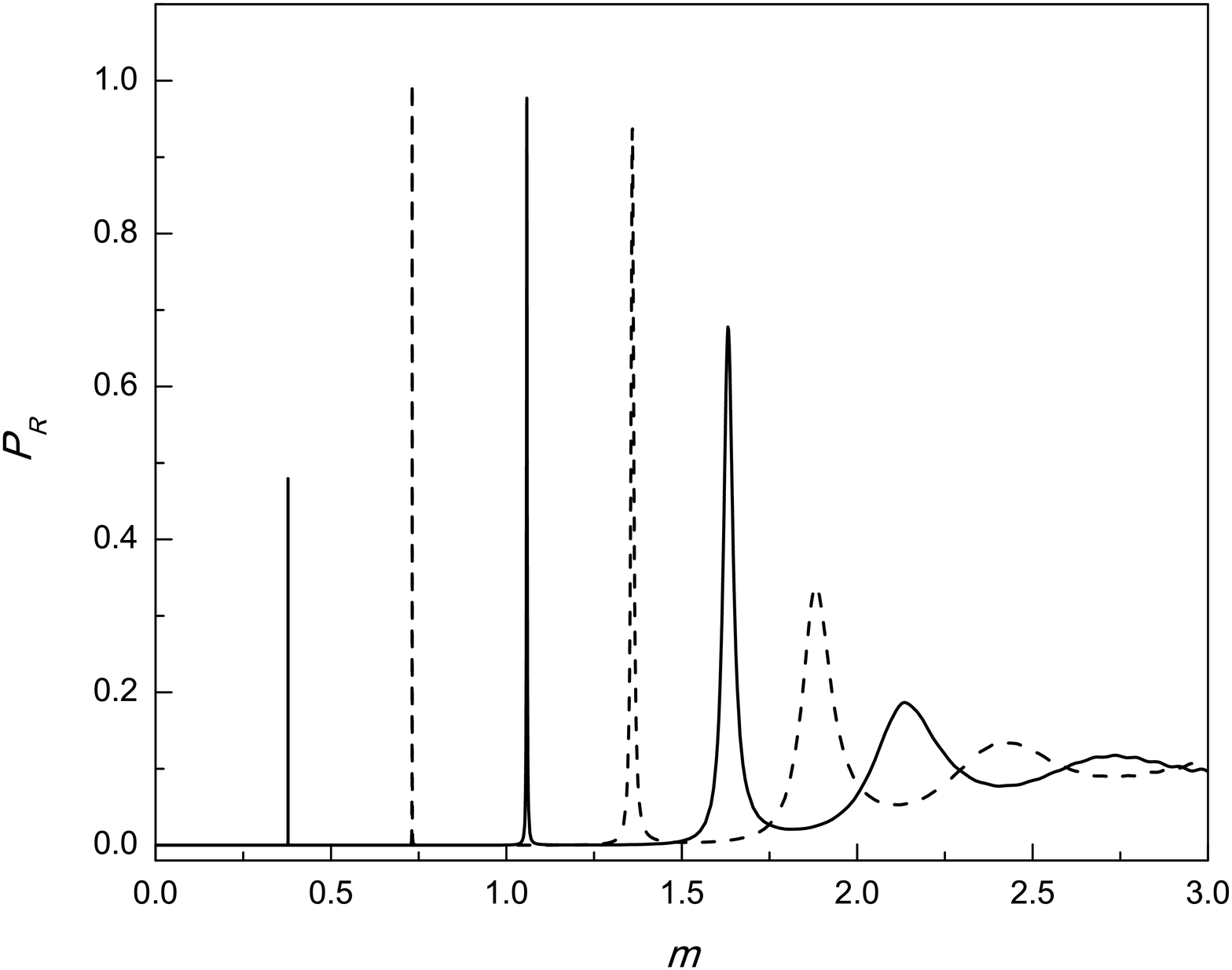}
\end{center}
\caption{\label{fig:mass-rightmodes} Plots of relative probability
for
 right handed fermion resonances with $a=1.0$(left column), $a=3.0$(middle column)
  and $a=5.0$(right column). Coupling constants are $\eta=0.5$(first raw), $\eta=1.0$(second raw)
  and $\eta=2.0$(third raw). Solid lines and dashed lines corresponds to even and odd parity respectively. In all cases $\lambda=1.0$.}
\end{figure*}
\begin{figure*}[htb]
\begin{center}
\includegraphics[width=4cm]{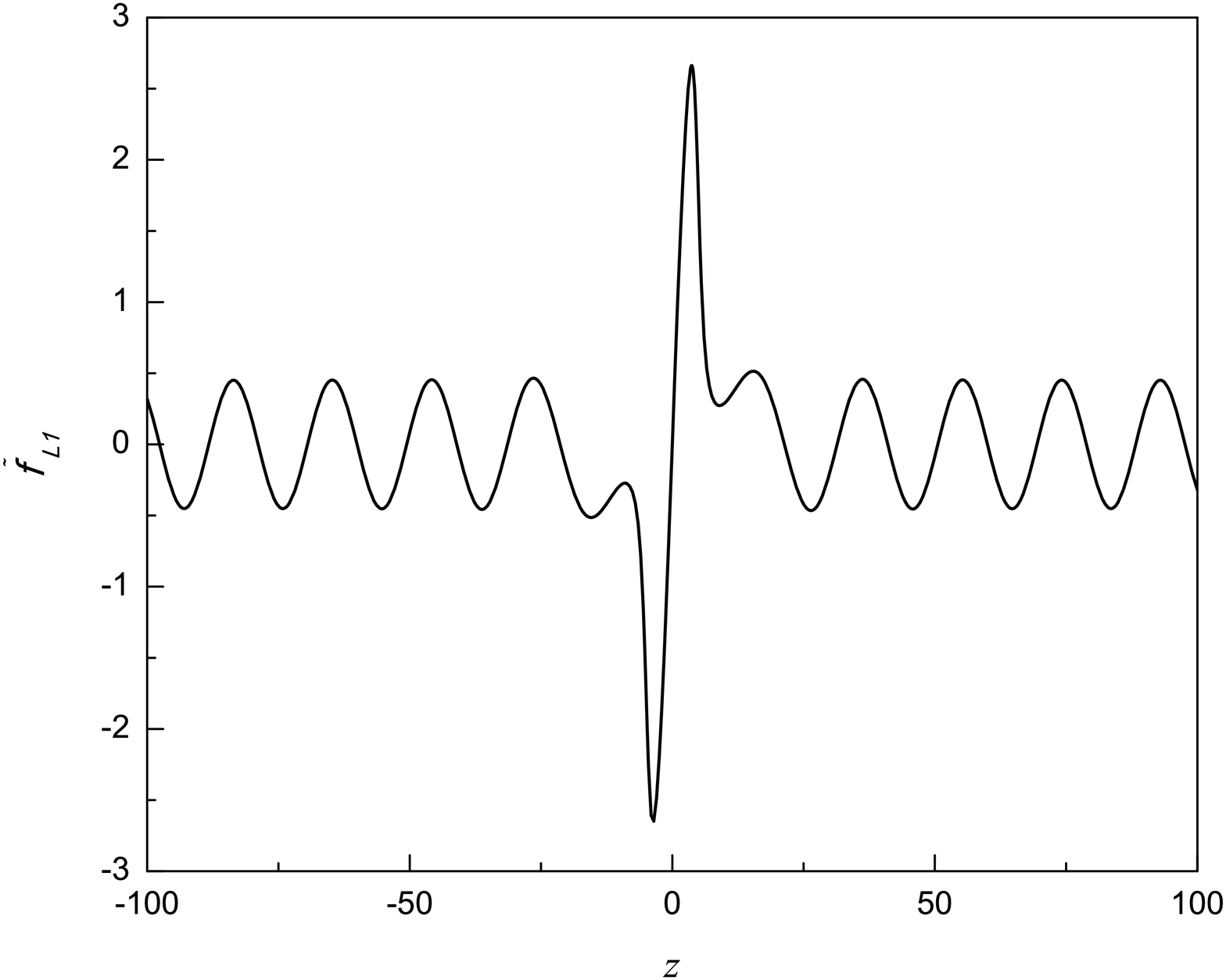}
\includegraphics[width=4cm]{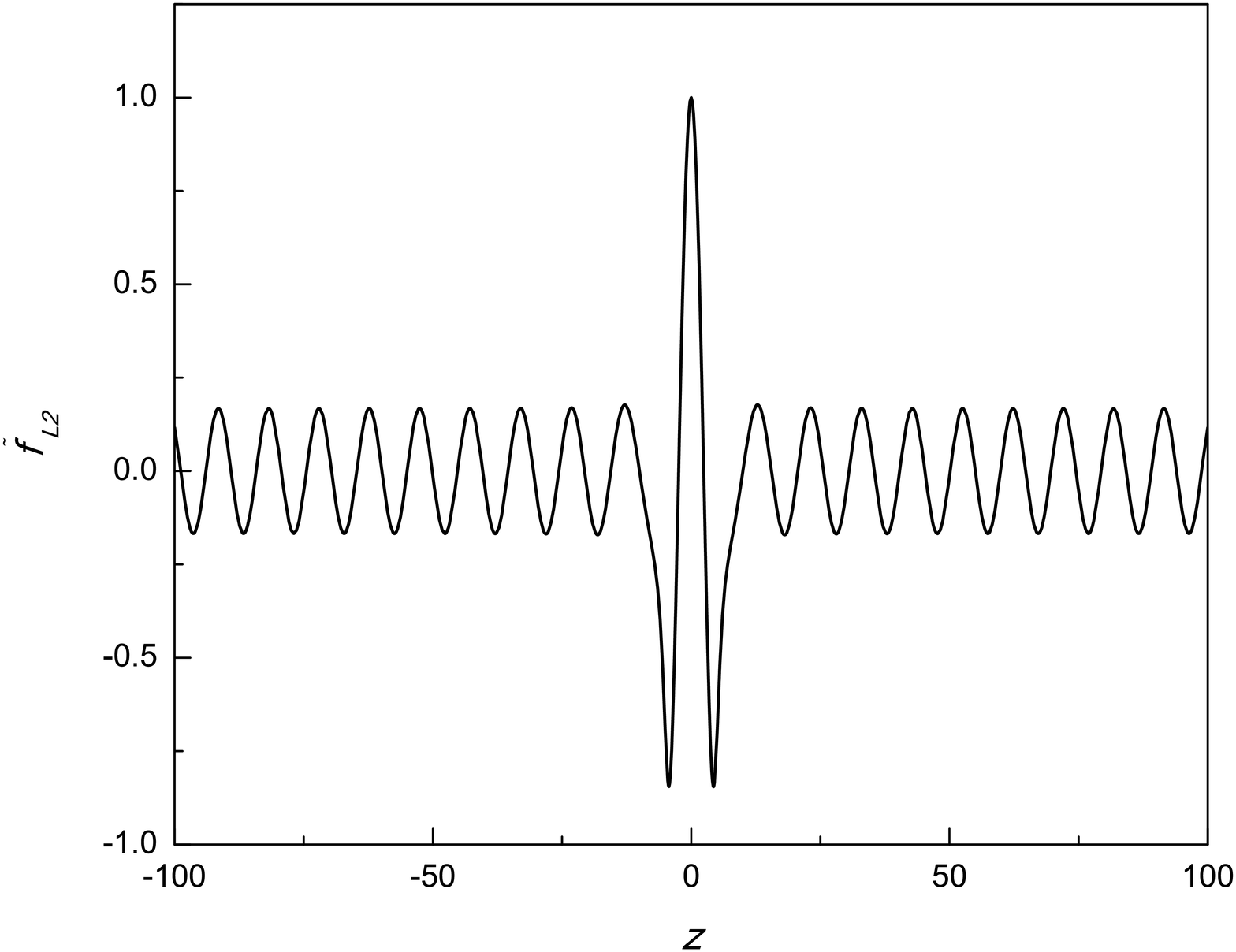}
\includegraphics[width=4cm]{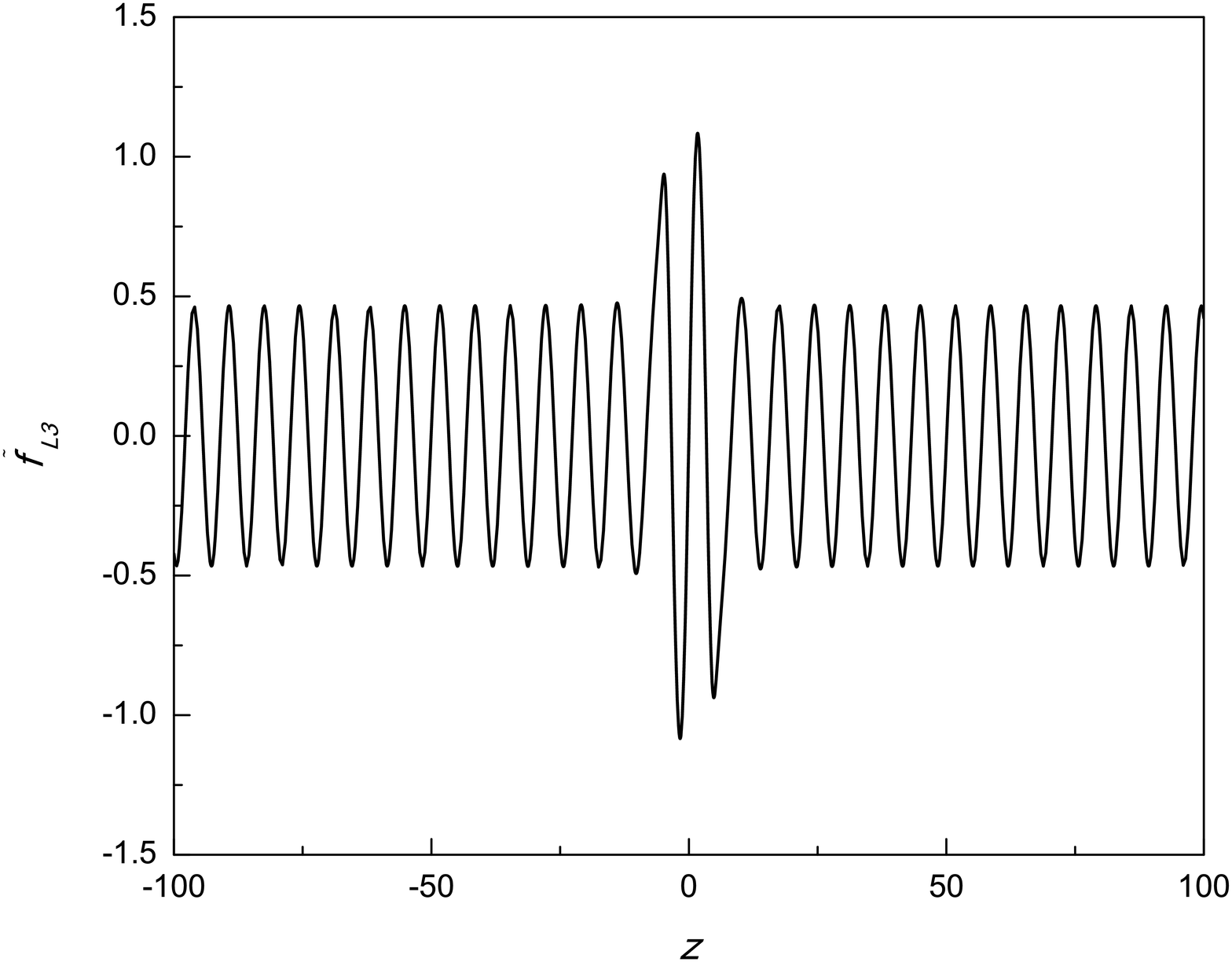}
\includegraphics[width=4cm]{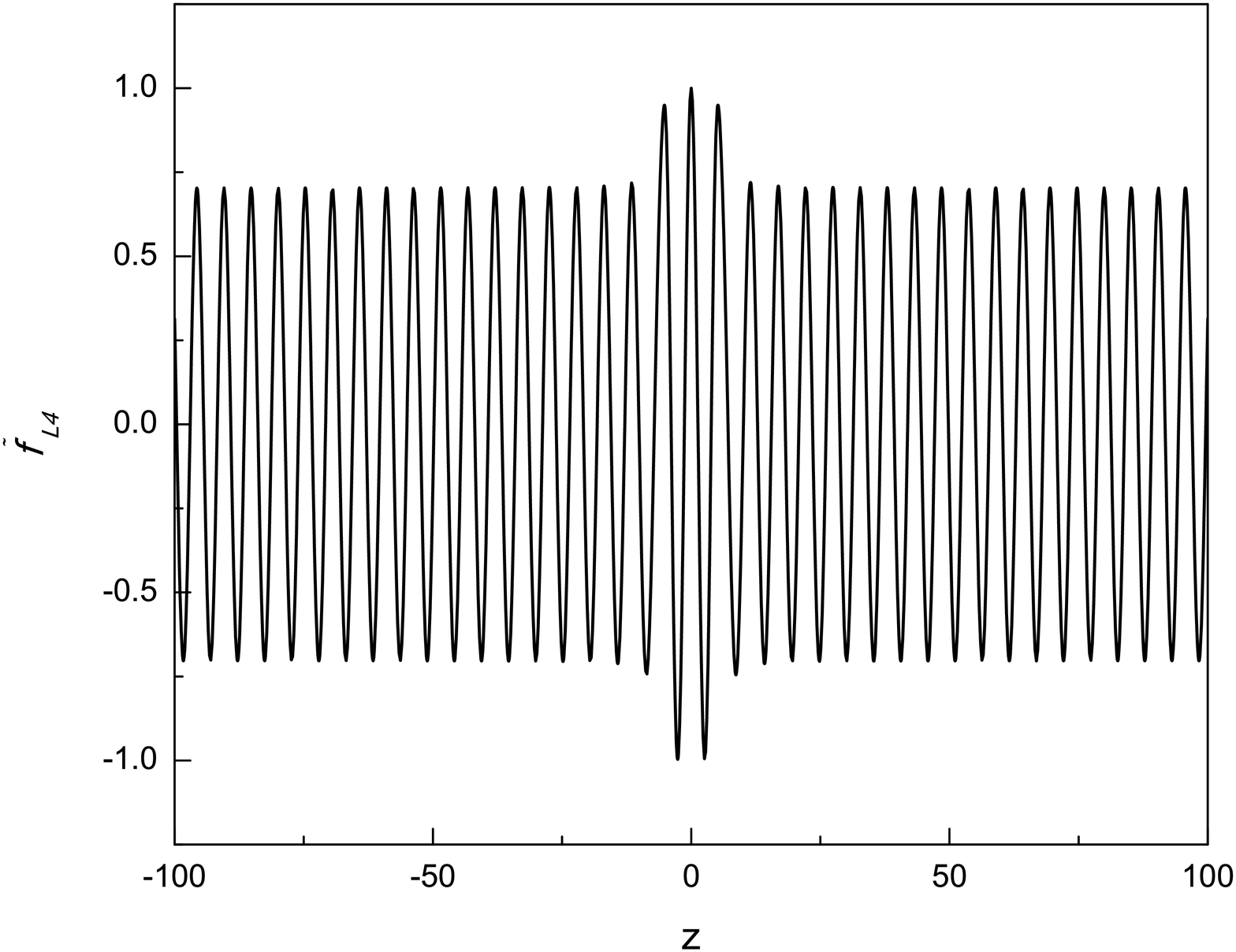}
\includegraphics[width=4cm]{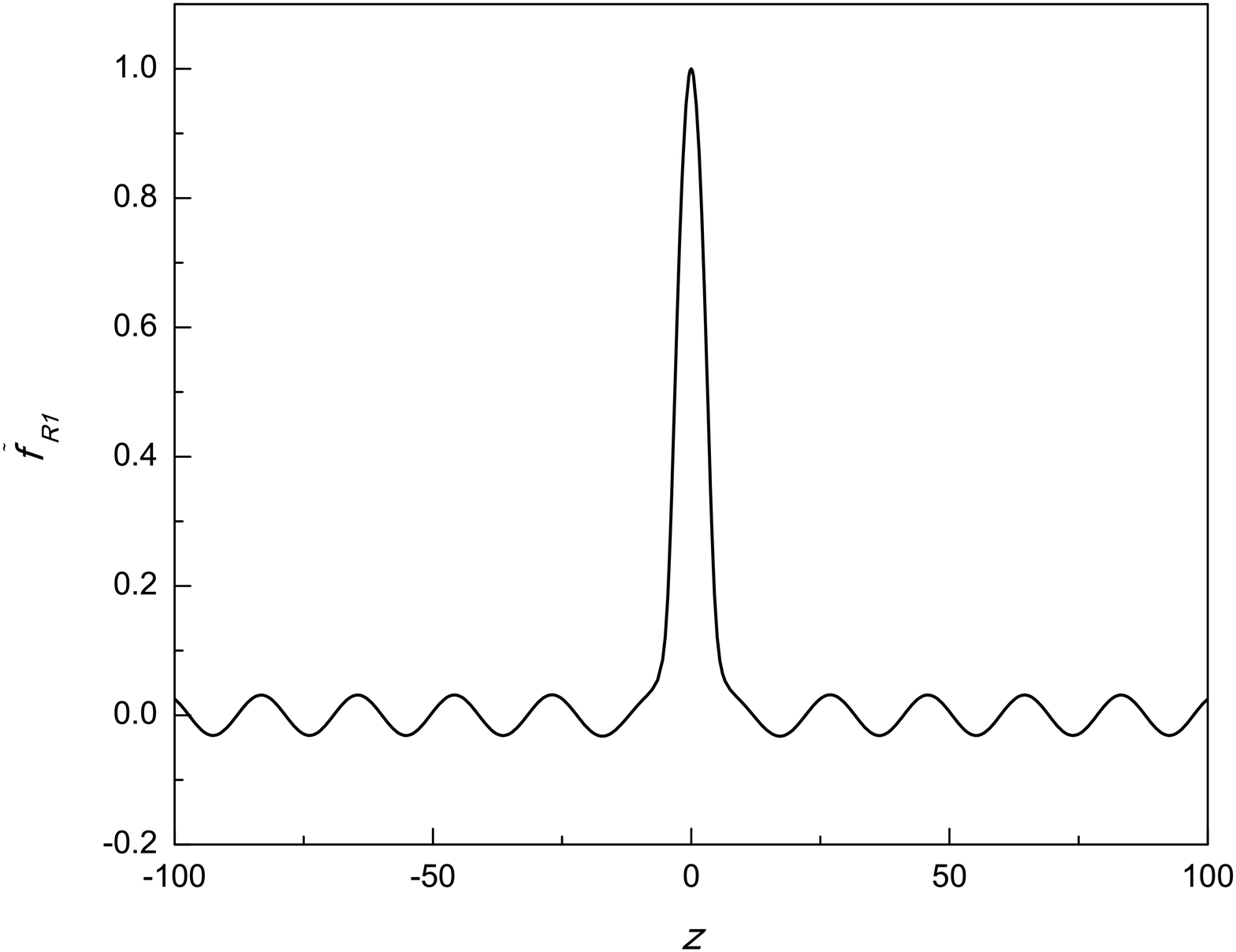}
\includegraphics[width=4cm]{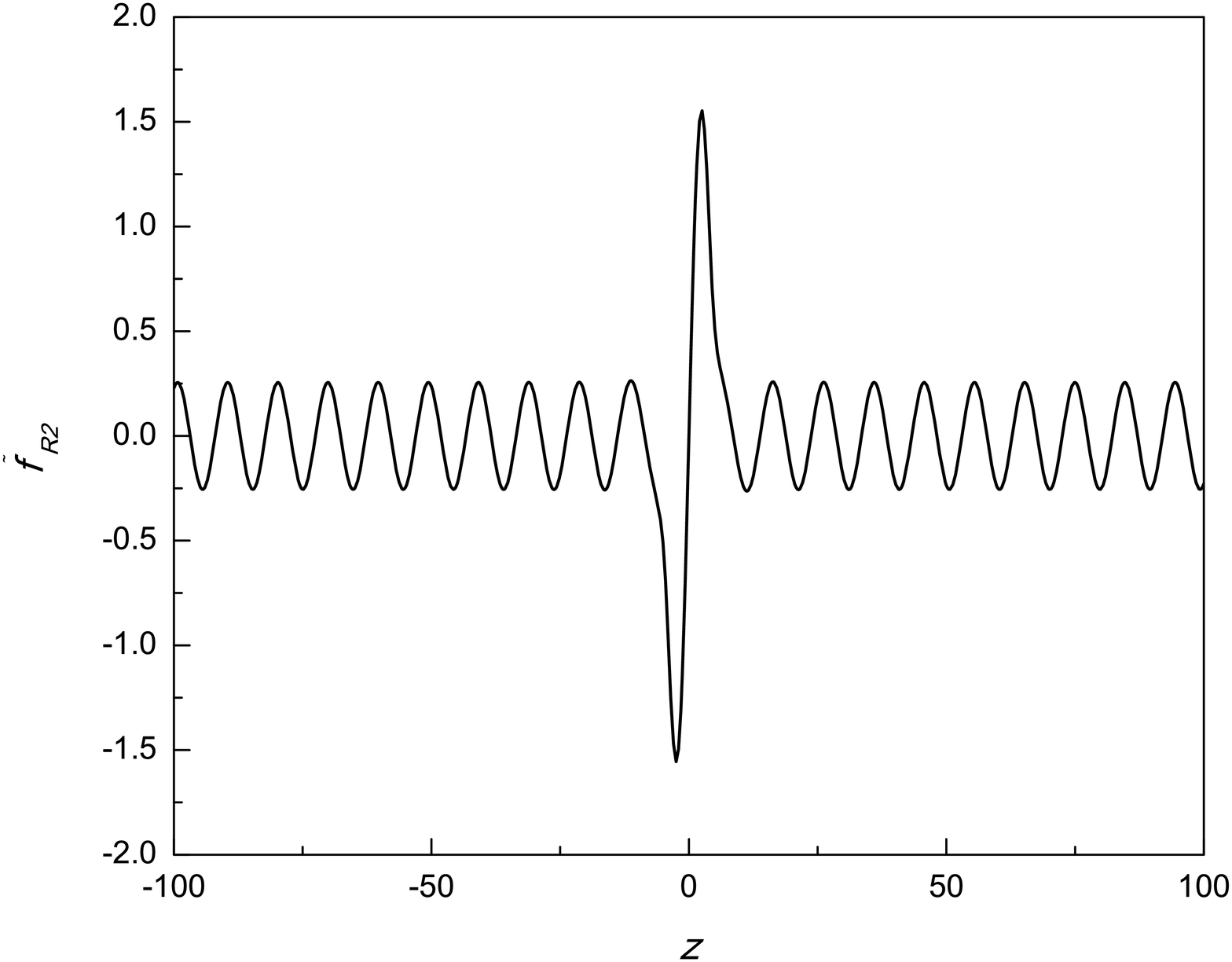}
\includegraphics[width=4cm]{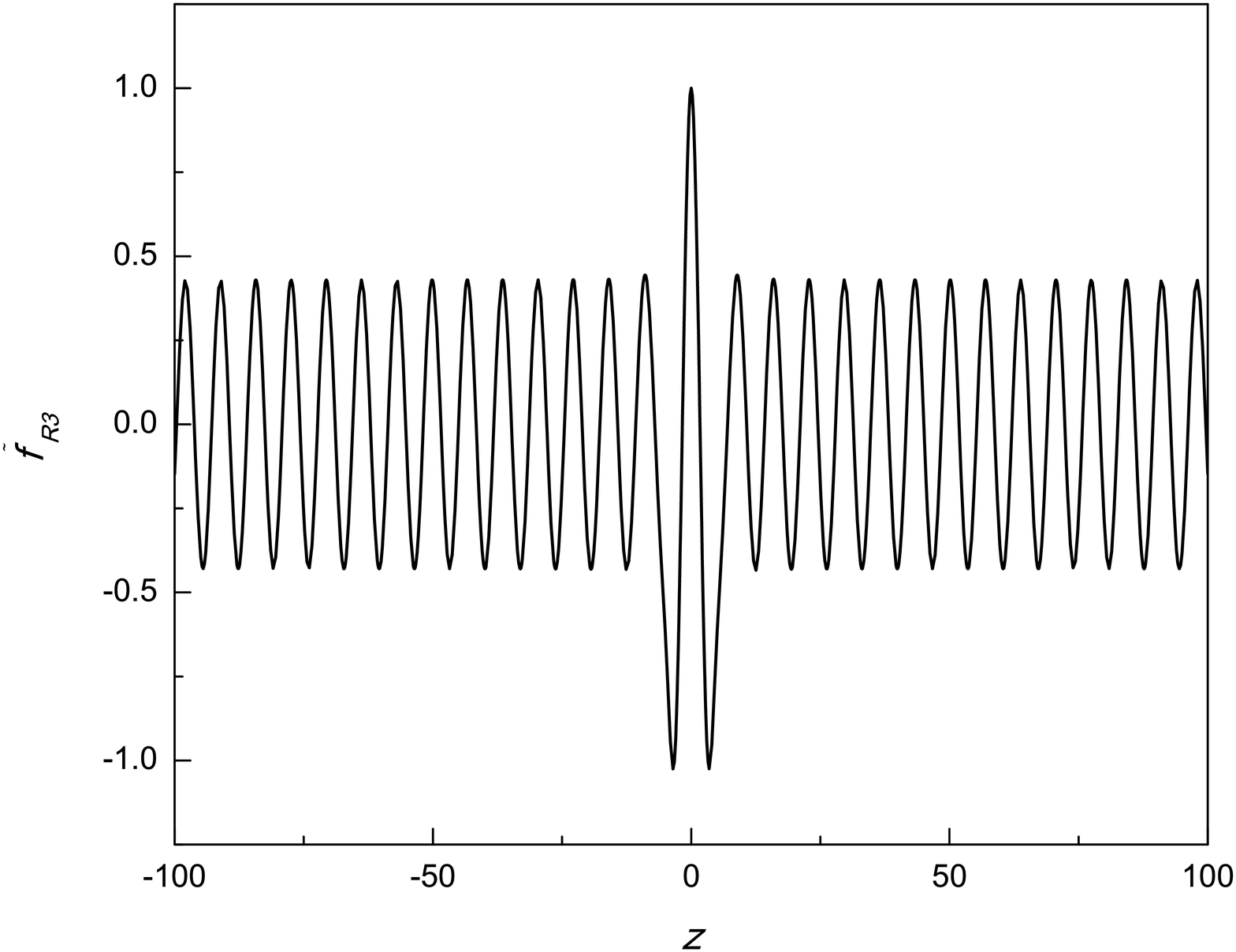}
\includegraphics[width=4cm]{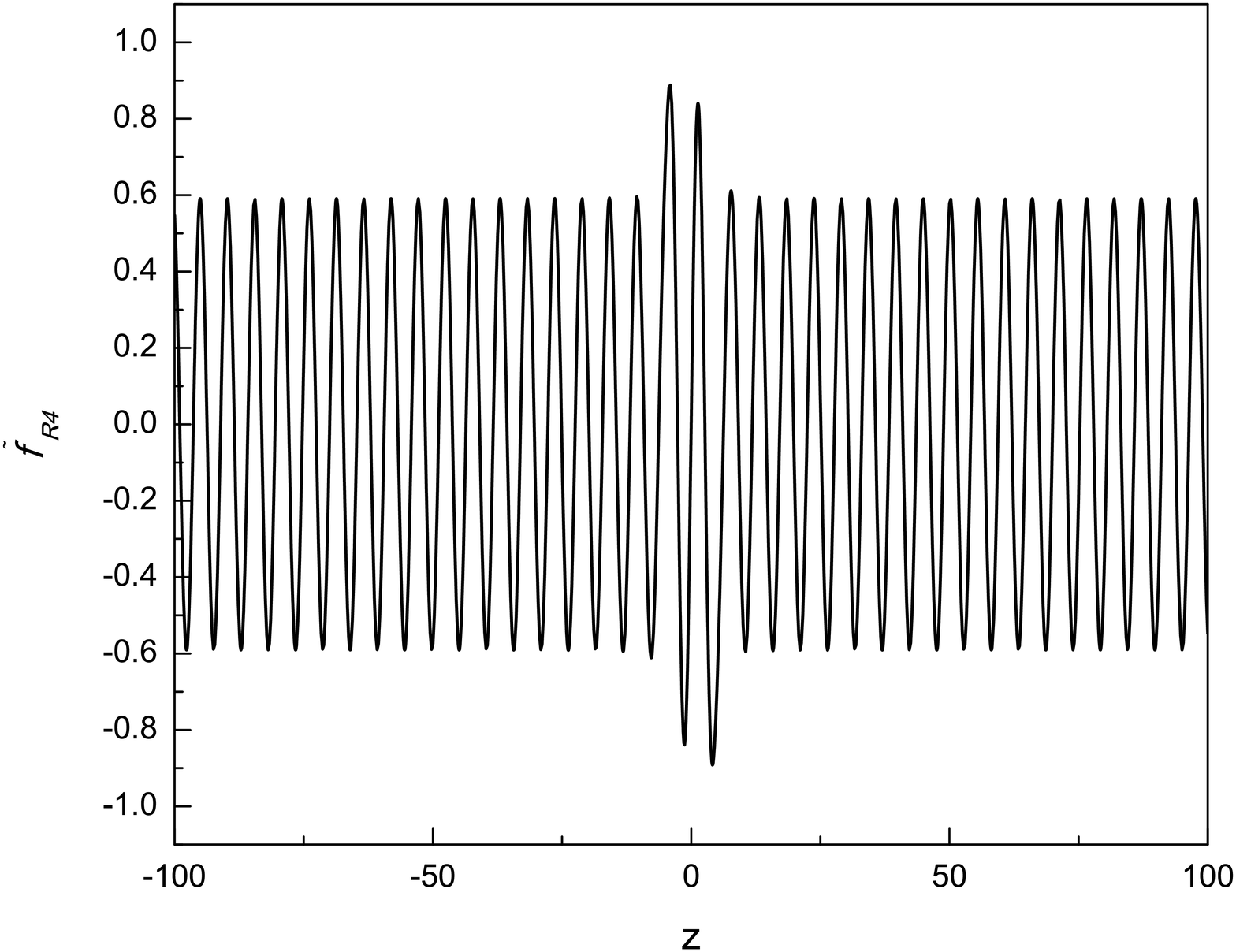}
\end{center}
\caption{\label{fig:resonances} Plots of the left-handed and
right-handed fermion resonances with $\eta =1$, $\lambda =1$ and
$a=5$.}
\end{figure*}
The asymptotic behaviors of the potentials are
\begin{eqnarray}
\label{eq:VLRasym}
&&V_{L}(0) = \eta \Big\{ \eta (\beta -1)^{2}-\big[\eta (\beta -1)^{2}+\lambda (\beta^{2}+1)\big] \sec\!h^{2}(\lambda a)\Big\} \\
&&V_{R}(0) = \eta \Big\{ \eta (\beta -1)^{2}-\big[\eta (\beta -1)^{2}-\lambda (\beta^{2}+1)\big] \sec\!h^{2}(\lambda a)\Big\} \\
&&V_{L}(\pm \infty) =V_{R}(\pm \infty)=0
\end{eqnarray}
It is realized that $V_{L}(0)$ can be negative or positive, while for $\eta>0$ and $\lambda>0$, $V_{R}(0)$
is always greater than $V_{L}(0)$.
\section{Resonances of massive modes}
\label{sec:massive-localize}

We can rewrite eqs. \eqref{eq:ScheqL} and \eqref{eq:ScheqR} as
\begin{equation}
\label{eq:supersym}
\begin{split}
Q^{\dagger}Q\tilde{f}_{Ln}=(-\partial_{z}+\eta Fe^{A})(\partial_{z}+\eta Fe^{A})\tilde{f}_{Ln}=m^{2}\tilde{f}_{Ln}
\\
QQ^{\dagger}\tilde{f}_{Rn}=(\partial_{z}+\eta Fe^{A})(-\partial_{z}+\eta Fe^{A})\tilde{f}_{Rn}=m^{2}\tilde{f}_{Rn}
\end{split}
\end{equation}
From these equation, we can see that the tachyonic modes in spectrum
excluded. With converting the equations of motion for fermion to
Schr\"{o}dinger-like equations, we can present a quantum mechanical
interpretation for $\tilde{f}_{Ln}$ and $\tilde{f}_{Rn}$. By
studying resonant modes we are able to obtain information about the
coupling between massive modes and the brane.

In order to derive KK modes from equation \eqref{eq:ScheqL} and \eqref{eq:ScheqR}
 we apply relative probability method \cite{Almeida:2009jc,Liu:2009dw,Liu:2009ve}.
  Since the equations \eqref{eq:ScheqL} and \eqref{eq:ScheqR} are Schr\"{o}dinger-like,
  we can interpret normalized $|\tilde{f}_{L,R}(z)|^{2}$ as the propability of finding massive KK modes on the brane.
  But the massive modes can not be normalized because they are oscillating when
   far away from brane along extra dimension. Therefore, the relative probability function
   is defined as \cite{Almeida:2009jc}
\begin{equation}
P_{Ln,Rn}(m)=\frac{\int_{-z_{b}}^{z_{b}}dz|\tilde{f}_{Ln,Rn}(z)|^{2}}{\int_{-z_{max}}^{z_{max}}dz|\tilde{f}_{Ln,Rn}(z)|^{2}}
\label{eq:PLR}
\end{equation}
where $2z_{b}$ is brane thickness approximately and $z_{max}=10z_{b}$ here. $\tilde{f}_{Ln,Rn}(z)$ is
solution of eq. \eqref{eq:ScheqL} or \eqref{eq:ScheqR} with two boundary conditions:
\begin{equation}
\tilde{f}_{Ln,Rn}(0)=1,
\qquad
\tilde{f}_{Ln,Rn}^{\prime}(0)=0,
\label{eq:inicon1}
\end{equation}
for even parity, and
\begin{equation}
\tilde{f}_{Ln,Rn}(0)=0,
\qquad
\tilde{f}_{Ln,Rn}^{\prime}(0)=1
\label{eq:inicon2}
\end{equation}
for odd parity. If $m^{2}\gg V_{L,R max}$, $\tilde{f}_{L,R}$ will be approximately a
 plane wave with $P_{L,R}(m)=z_{b}/z_{max}=0.1$. For simplicity
 we concern $\beta=1$. Figs. \ref{fig:mass-leftmodes} and \ref{fig:mass-rightmodes}
 show plots of the relative probability for different values of $a$ and $\eta$ for left-handed and
  right-handed fermions respectively. The masses of resonances were presented in Table 1. For simplicity the
  fermion resonant wavefunctions for $\eta=1$, $\lambda=1$ and $a=5.0$ were showed in Fig. \ref {fig:resonances}.
\\
From the figures, we can see that the spectra of massive KK modes of left handed
 and right handed fermions are almost the same which reveals that a Dirac fermion
 composed from left handed and right handed bound KK modes. Furthermore for
  fixed value of $a$ the larger value of coupling parameter, the larger number
  of resonances. Also for fixed value of $\eta$, the lager value of $a$ leads to
  larger number of peaks. The first peak is the most narrow and the resonances
  will become broader with increasing $m$. This means that first resonance has
  larger lifetime and the lifetime decline with increasing the mass of peak.
  \\
  We can also see from figures that there are successively even and odd parity wavefunctions
   for left and right chiral modes with the same values of $m^2$. In other words, the zero mode
    is begining of left chirality with even parity spectrums, therefore the two first resonances (if exist)
    with the same $m^{2}$ are odd parity left chiral mode and even parity right chiral mode. Next
    the two second resonances (if exist) with the same $m^{2}$ are even parity left chiral mode
    and odd parity right chiral mode.
\begin{table}[htbp]
\centering         \begin{scriptsize}
        \begin{tabular}{|c|c|p{2.5cm}|p{5cm}|}
        \hline {Left} & $\eta=0.5$ & $\eta=1.0$ & $\eta=2.0$  \\
        \hline $a=1.0$
 & absent & 1.185465 & 1.719258 \\
        \hline $a=3.0$ & 0.453441 & 0.562819;0.994057 & 0.689911;1.232968;1.664278;
        2.034932  \\
        \hline$a=5.0$
 & 0.288833;0.567862 & 0.337121;0.646626;
 0.924378;1.198549
 & 0.377447;0.731511;1.058468;1.359395;
 1.631559;1.884065;2.146727\\
      \hline  \hline{Right}  & $\eta=0.5$ & $\eta=1.0$ & $\eta=2.0$  \\
        \hline $a=1.0$
 & absent & 1.110830 &  1.711422 \\
        \hline $a=3.0$ & 0.442279 & 0.562819;0.994057 &  0.689911;1.232968;1.662400;2.027853 \\
        \hline $a=5.0$
 & 0.283917;0.561504 & 0.337121;0.645689;
 0.921857;1.190822
 &0.377447;0.731480;1.058468;1.359395;
  1.631559;1.882145;2.139101\\
        \hline
        \end{tabular}
               \caption{Masses of resonances for different values of $\eta$ and $a$.}  \end{scriptsize}
   \end{table}
\section{Conclusions}
In this paper we have studied the issue of the localization of
fermion field on the double wall brane. This brane includes two
scalar field coupled minimally to brane. For observing weather the
zero mode can be localized on this brane or not, we use a Yukawa
coupling between fermion and background scalar fields. By
investigating normalization condition, we can see that fermion
couples with summation of kinks . we found that there is a relation
between coupling constant and $\lambda$ parameter. Also we found
that the zero mode of right handed fermion can not be localized on
the brane.
\\
The massive mode resonances were investigated numerically. From the
volcano shape of effective potential for left handed fermions, it
results that the spectrum is continuous and there is no gap between
zero mode and KK excitation modes. Also larger values of coupling
constant and the distance of sub-branes support more resonances in
the spectrum. In one spectrum, heavier resonances have broader peaks
rather than lighter ones. This means the lighter fermions couple
stronger to the brane rather than heavier KK modes. because of very
narrow peak, We may not see light resonances. Fortunately because of
supersymmetric feature of Schr\"{o}dinger-like equation, we have successively
even and odd parity for left or right chirality modes with the same mass in the spectrum.
This helps us to search a small region when a resonance peak
is not seen. Therefore numeric procedure becomes simple and fast.
The lifetime of a resonance is proportional to inverse of peak width
at half maximum. Hence generally, light resonances have longer
lifetimes.


\end{document}